\documentclass[a4paper,fleqn,usenatbib]{mnras}

\usepackage{newtxtext,newtxmath}

\usepackage[T1]{fontenc}
\usepackage{ae,aecompl}

\usepackage{graphicx}	
\usepackage{amsmath}	
\usepackage{amssymb}	
\usepackage{rotating}
\usepackage{lscape}
\usepackage{dblfloatfix}

\title[Spectral study of A523]{Spectral study of the diffuse synchrotron source in the galaxy cluster Abell 523}

\author[V. Vacca et al.]{Valentina Vacca,$^{1}$\thanks{E-mail: valentina.vacca@inaf.it}, Timothy Shimwell$^{2}$, Richard A. Perley$^{3}$, Federica Govoni$^{1}$, Matteo Murgia$^{1}$, 
\newauthor Luigina Feretti$^{4}$, Gabriele Giovannini$^{4,5}$, Francesca Loi$^{1}$, Ettore Carretti$^{4}$, Filippo Cova$^{6}$, 
\newauthor 
Fabio Gastaldello$^{6}$, 
Marisa Girardi$^{8}$, Torsten En{\ss}lin$^{7}$, Hiroki Akamatsu$^{9}$, Annalisa Bonafede$^{4,5}$, 
\newauthor 
Etienne Bonnassieux$^{5}$, Walter Boschin$^{10,11,12}$,
 Andrea Botteon$^{9}$, Gianfranco Brunetti$^{4}$, Marcus Br{\" u}ggen$^{13}$, 
\newauthor
Alexis Finoguenov$^{14}$, Duy Hoang$^{13}$, Marco Iacobelli$^{2}$, Emanuela Orr\'u$^{2}$, Rosita Paladino$^{4}$, Huub R{\"o}ttgering$^{9}$, 
\newauthor
Reinout van Weeren$^{9}$, Fabio Vitello$^{4}$, Denis Wittor$^{13,5}$\\
$^{1}$INAF-Osservatorio Astronomico di Cagliari, Via della Scienza 5, I-09047 Selargius (CA), Italy \\
$^{2}$ASTRON, the Netherlands Institute for Radio Astronomy, Postbus 2, 7990
AA, Dwingeloo, The Netherlands\\
$^{3}$National Radio Astronomy Observatory, P.O.Box O, Socorro, NM, 87801\\
$^{4}$INAF - Istituto di Radioastronomia, Via P. Gobetti 101, 40129 Bologna, Italy\\
$^{5}$Dipartimento di Fisica e Astronomia, Universita di Bologna, Via Gobetti 93/2, 40122, Bologna, Italy\\
$^{6}$IASF - Milano, INAF, Via Corti 12, I-20133 Milan, Italy\\
$^{7}$Max Planck Institute for Astrophysics, Karl-Schwarzschildstr. 1, 85741 Garching, Germany\\
$^{8}$Dipartimento di Fisica dell’Università degli Studi di Trieste - Sezione di Astronomia, via Tiepolo 11, I-34143 Trieste, Italy\\
$^{9}$Leiden Observatory, Leiden University, P.O. Box 9513, 2300 RA Leiden, The Netherlands\\
$^{10}$Fundaci\'on G. Galilei - INAF (Telescopio Nazionale Galileo), Rambla J. A. Fern\'andez P\'erez 7, E-38712 Bre$\widetilde{n}$a Baja (La Palma), Spain\\
$^{11}$Instituto de Astrof\'isica de Canarias, C/V\'ia Lactea s/n, E-38205 La Laguna (Tenerife), Spain\\
$^{12}$Departamento de Astrof\'isica, Univ. de La Laguna, Av. del Astrof\'isico Francisco S\'anchez, s/n, E-38205 La Laguna (Tenerife), Spain\\
$^{13}$University of Hamburg, Gojenbergsweg 112, 21029 Hamburg, Germany\\
$^{14}$Department of Physics, University of Helsinki, Gustaf Hällströmin katu 2, FI-00014 Helsinki, Finland}

\date{Accepted for publication on MNRAS}

\pubyear{}

\begin{document}
\label{firstpage}
\pagerange{\pageref{firstpage}--\pageref{lastpage}}
\maketitle

\begin{abstract}
The galaxy cluster Abell 523 (A523) hosts an extended diffuse synchrotron source historically classified as a radio halo. Its radio power at 1.4\,GHz makes it one of the most significant outliers in the scaling relations between observables derived from multi-wavelength observations of galaxy clusters: it has a morphology that is different and offset from the thermal gas, and it has polarized emission at 1.4\,GHz typically difficult to observe for this class of sources. A magnetic field fluctuating on large spatial scales ($\sim$1\,Mpc) can explain these peculiarities but the formation mechanism for this source is not yet completely clear. To investigate its formation mechanism, we present new observations obtained with the LOw Frequency ARray at 120-168\,MHz and the Jansky Very Large Array at 1-2\,GHz, which allow us to study the spectral index distribution of this source. According to our data the source is observed to be more extended at 144\,MHz than previously inferred at 1.4\,GHz, with a total size of about 1.8\,Mpc and a flux density $S_{\rm 144\,MHz}=(1.52\pm0.31)$\,Jy. The spectral index distribution of the source is patchy with an average spectral index $\alpha\sim1.2$ between 144\,MHz and 1.410\,GHz, while an integrated spectral index $\alpha\,\sim\,2.1$ has been obtained between 1.410\,GHz and 1.782\,GHz.
A previously unseen patch of steep spectrum emission is clearly detected at 144\,MHz in the south of the cluster. 
Overall, our findings suggest that we are observing an overlapping of different structures, powered by the turbulence associated with the primary and a possible secondary merger.

\end{abstract}





\section{Introduction}
\label{intro}
Diffuse synchrotron sources observed in merging galaxy clusters indicate the presence of ultra-relativistic electrons ($\gamma>>1000$) spiralling along the flux lines of weak magnetic fields ($\sim\mu$G), and can be classified as (giant) halos and relics \citep[e.g.,][]{Feretti2012,Brunetti2014,Vacca2018a,vanWeeren2019}. 

Radio halos have been observed in a fraction of massive and disturbed
galaxy clusters where they fill the central volume of the system
\citep{Cassano2010b}. The prototype and most famous radio halo is Coma
C which resides in the Coma cluster and was first observed by
\cite{Large1959}. Overall, radio halos extend up to spatial scales of
$\sim$1-2\,Mpc and are characterized by low radio brightness
($\sim$0.1\,$\mu$Jy/arcsec$^2$ at 1.4\,GHz). Out of about 100 radio
halos presently confirmed, three show filaments of polarized emission
at around 20\% at 1.4\,GHz (A2255 \citealt{Govoni2005},
MACS\,J0717.5+3745 \citealt{Bonafede2009}, and A523
\citealt{Girardi2016}).  These are likely projected on the cluster
center as suggested by recent studies of A2255 and MACS\,J0717.5+3745
\citep{Pizzo2011,Botteon2020a,Rajpurohit2020}. Moreover, a 5$\sigma$
upper limit of 13\% on the fractional polarization has been derived
for the radio halo in 1E\,0657-55.8 by
\cite{Shimwell2014}. Three-dimensional numerical simulations suggest
that radio halos are intrinsically polarized at 1.4\,GHz, with
filaments of polarized emission expected to be observed at distances
greater than 1.5\,Mpc from the cluster center, reflecting the
intracluster magnetic field structure \citep{Loi2019}. However, the
resolution and sensitivity of present instruments hinder the detection
of this polarized emission in most radio halos \citep{Govoni2013}.

Typically, most powerful and extended radio halos are hosted in massive and bright X-ray clusters and show strong correlation between the integrated radio halo emission at 1.4\,GHz and the X-ray luminosity of the intracluster medium (ICM) and galaxy cluster mass \citep[e.g.,][]{Liang2000,Cassano2013,Yuan2015,Cuciti2021}. 
Exceptions have been observed of radio halos that show a radio power at 1.4\,GHz larger than expected from the radio power ($P_{\rm 1.4\,GHz}$) - X-ray luminosity ($L_{\rm X, 0.1-2.4\,keV}$) correlation holding for the most of radio halos, one of the most significant being the source in A523 \citep{Giovannini2011}\footnote{The other two galaxy clusters found to be significant outliers are A1213 \citep{Giovannini2009} and CLG\,0217+70 \citep{Brown2011b}. \cite{Zhang2020} recently demonstrate that by adopting a new revised redshift estimate, the galaxy cluster CLG\,0217+70 is not anymore overluminous in radio compared to X-ray, while the case of A1213 is still under investigation.}. \,Detailed studies indicate, in some cases, a point-to-point correlation between the X-ray and radio brightness at 1.4\,GHz as, e.g., for the radio halos in the Coma cluster and  1RXS\, J0603.3+4214 \citep{Govoni2001a,Brown2011a,Rajpurohit2018}, while a minor (e.g., A520 \citealt{Govoni2001b,Hoang2019}) or absent (e.g., 1E\,0657-55.8 \citealt{Shimwell2014}) correlation has been found in other clusters. 

Radio relics are elongated diffuse synchrotron sources with filamentary morphology that are observed in the periphery of a number of merging galaxy clusters. Sometimes they come in pairs along the merger axis but on opposite sides  \citep[e.g.,][]{Rottgering1997,Bagchi2006,vanWeeren2011} and/or coexist with cluster radio halos \citep[e.g.,][]{Brown2011b,Lindner2014,Parekh2017}. They are extended on spatial scales of $\sim$0.5-2\,Mpc, are characterized by low radio surface brightness ($\sim$0.1\,$\mu$Jy/arcsec$^2$ at 1.4\,GHz), and are highly polarized at GHz frequencies with fractional polarization larger than 20\% at 1.4\,GHz. 

In our present understanding, a key role in the origin of radio halos and relics is likely played by cluster merger phenomena in the form of turbulence and shocks.
Merger-induced turbulence is thought to accelerate a pre-existing electron population (seed population) through the Fermi-II mechanism, giving rise to diffuse synchrotron sources filling the central volume, i.e. radio halos, of a fraction of massive galaxy clusters. According to this scenario, a variety of spectra and spectral trends are expected \citep[see, e.g.][]{Brunetti2014}. Compelling evidence is now available that directly links radio relics with cluster shocks observed at X-rays and mm/sub-mm wavelengths \citep[e.g., ][]{Akamatsu2013,Planck2013,Erler2015,Eckert2016,Urdampilleta2018}. According to these findings, the emission observed in radio seems to be explained if cosmic-ray electrons are accelerated up to $\sim$GeV energies through diffusive shock acceleration \citep[DSA][]{Drury1983,Blandford1978} either from the thermal pool, as proposed first by \cite{Ensslin1998}, or from supra-thermal/relativistic plasma (e.g., \citealt{Markevitch2005} and \citealt{Kang2012}).

Precious insights on the origin of these sources come from the spectral analysis of their properties over a wide frequency range and from the comparison of observations with theoretical expectations and simulations. Radio halos are characterized by steep-spectra ($S_{\nu}\propto \nu^{-\alpha}$, with $\alpha\simeq$1-1.4, where $S_{\nu}$ is the flux density at the observing frequency $\nu$ and $\alpha$ is the spectral index). Detailed spatially-resolved spectral index images have been produced for a number 
of radio halos, revealing different behaviours. Indications of radial steepening have been found in a few clusters \citep[e.g.,][]{Giovannini1993,Feretti2004,Pearce2017}, while other radio halos show no clear trend with either  uniform \citep[e.g.,][]{Vacca2014,Rajpurohit2018,Hoang2019}
or complex \citep[e.g.,][]{Giacintucci2005,Orru2007,Kale2010,Shimwell2014,Rajpurohit2020,Botteon2020a} distributions. Radio relics are also characterized by steep-spectra ($\alpha\simeq$1-1.5), and detailed spectral index images show a steepening transversely to their elongation towards the cluster center \citep[e.g., ][]{Clarke2006,deGasperin2015,diGennaro2018}.  The steepening is interpreted as synchrotron and inverse Compton losses  in  the  shock  downstream region. 

In this paper we present new 144\,MHz LOw Frequency ARray (LOFAR) observations and new 1-2\,GHz Jansky Very Large Array (VLA) observations of the galaxy cluster A523 which hosts a powerful diffuse source associated with the ICM,   historically classified as a radio halo. Its radio power at 1.4\,GHz makes it one of the most significant outliers in the scaling relations between observables derived from multi-wavelength observations of galaxy clusters: it has a morphology that is different and offset from the thermal gas, and it has polarized emission at 1.4\,GHz typically difficult to observe for this class of sources. 
We study the spectral behaviour of the emission between 144\,MHz and 2\,GHz to better understand the formation mechanism of this source and compare the radio emission with observations at X-ray wavelengths. 
In \S\ref{A523} we summarize the present knowledge on this cluster, in \S\ref{observations} we describe the data used for the analysis. In \S\ref{lowfreq} we present the new images of the cluster, in \S\ref{spix_sect} the spectral behaviour and in \S\ref{Radio_versus_X} a comparison between the radio and X-ray derived cluster properties. Finally, in \S\ref{discussion} and in \S\ref{conclusions} we discuss the nature of the system and draw our conclusions. Throughout, we use a $\Lambda$CDM cosmology with $H_{0}=67.4$\,km/s/Mpc, $\Omega_0=0.315$ and $\Omega_{\Lambda}=0.685$ \citep{Planck2020}. With this cosmology, at the distance of A523 ($z=0.104$, luminosity distance 499\,Mpc), $1^{\prime\prime}$ corresponds to 1.98\,kpc. 

\begin{table*}
\caption{Radio observations used in this work.}
        \begin{tabular}{ccccccccc}
         \hline
          \hline
 RA & Dec& Instrument &$\nu$& Bandwidth         &Config.      & Date & Duration&Project \\
h:m:s&$^{\circ}$:$^{\prime}$:$^{\prime\prime}$&& (MHz) & (MHz)&& & (h) &\\
(J2000) &(J2000) & & && &  &\\
\hline\hline
04:57:06.3&+08:56:53.7&LOFAR&144&48&HBA DUAL &31-10-2018 & 4& LC10$\_$024    \\
04:57:06.3&+08:56:53.7&LOFAR&144&48&HBA DUAL &11-10-2018 & 4& LC10$\_$024    \\
\hline
04:59:10.0    &  +08:49:00.0& VLA&1.5& 1000& C& 06-07-2013&1.1& 13A-168\\
04:59:10.0    &  +08:49:00.0& VLA&1.5& 1000& D& 29-01-2013&1.1& 13A-168\\

                  \hline
            \hline
        \end{tabular}
        \label{tab:B}

\end{table*}
\section{The galaxy cluster A523}
\label{A523}
A523 is a nearby cluster ($z=0.1040\pm0.0004$, \citealt{Girardi2016})
with an estimated $M_{\rm X, 500}\approx 2.2-3.6 \times 10^{14}M_{\odot}$
\citealt{Cova2019}, where $M_{\rm X, 500}$ is the mass derived from X-observables
within $R_{\rm 500}$\footnote{We refer to $R_{\Delta}$
  as the radius of a sphere within which the mean mass density is $\Delta$ times the critical density at the redshift of the galaxy cluster system. $M_{\Delta}$ is the mass contained in $R_{\Delta}$.}. 
A523 is a disturbed system with an ongoing merger along the SSW-NNE direction
between two subclusters \citep{Girardi2016,Golovich2019}. A secondary
merger is likely present along the ESE-WNW direction
\citep{Cova2019}. The cluster is known to host a powerful extended
synchrotron source, classified as a radio halo because it permeates
both the merging clumps and does not show transverse flux asymmetry
typical of radio relics \citep{Giovannini2011}.  \cite{Girardi2016}
derived a flux density of the source at 1.4\,GHz of
$72\pm3$\,mJy. According to \cite{Einasto2001}, A523 belongs to the
supercluster SCL\,62 that includes the clusters A525, A515, A529,
A532, all of which lie at a similar redshift. In this context, the
supercluster has been recently targeted to search for diffuse emission
in and beyond galaxy clusters with the Sardinia Radio Telescope (SRT)
by \cite{Vacca2018} who found patches of diffuse emission of unknown
origin. A fraction of these turned out to be Galactic H-$\alpha$
regions and blending of discrete sources \citep{Hodgson2020}.  Follow
up studies are still in progress for the remaining regions.

A detailed multi-wavelength study including optical, X-ray, and radio
data at 1.4\,GHz has enabled an accurate characterization of the
cluster dynamical and non-thermal properties, revealing that the radio
halo emission is mainly elongated perpendicular to the merging axis
defined by the optical and X-ray observations, with a clear offset
between the peaks of the radio and the X-ray brightness
distribution of about 310\,kpc \citep{Girardi2016}. Unlike other
radio halos, the source in A523 does not show a strong point-to-point
correlation between the 1.4\,GHz radio and X-ray brightness
distribution but it is rather characterized by a broad intrinsic
scatter between these two quantities \citep{Cova2019}. Overall, its
radio power\footnote{k-correction was applied with a spectral
    index $\alpha=1.2$.} at 1.4\,GHz is $P_{\rm 1.4\,GHz}=2.2\times
10^{24}$\,W/Hz, while the X-ray luminosity is $L_{\rm X, 500
    [0.1-2.4]}=1.5\times 10^{44}$\,erg/s \citep{Girardi2016,Cova2019}. The 1.4\,GHz radio emission appears
  to be higher by a factor $\sim 6-18$ than expected from the X-ray
  luminosity of the system according to the best-fit scaling relations
  presented in \cite{Yuan2015}. We use for the redshift of this
  cluster $z=0.1040\pm0.0004$ \citep{Girardi2016}, derived through a
  detailed optical study based on 132 spectroscopic redshifts of which
  80 related to cluster members and later confirmed by
  \cite{Golovich2019}. This makes it very unlikely that an incorrect
  estimate of the redshift could be the reason of the higher radio
  power.

It might be that the cluster emission is overly bright in the radio
because of a third subcluster merging in the direction perpendicular
to the main merger and causing additional turbulence, as proposed by
\cite{Cova2019} on the basis of new \emph{XMM-Newton} and
\emph{Nustar} observations. Other clusters undergoing multiple mergers
have been found to host radio halos, but these radio halos are not
over-luminous in radio with respect to X-rays \citep[e.g.,
][]{Bonafede2009}. The peculiarity of A523 is apparent also in its
location on the $P_{\rm 1.4\,GHz}$ - $M_{\rm X, 500}$ correlation.
By considering the $P_{\rm 1.4\,GHz}$-$M_{\rm X, 500}$
  correlation for radio halos found by \cite{Yuan2015}, we derive a
  mass $M_{\rm X, 500}\,\approx\,6-10\,\times\,10^{14}\,M_{\odot}$,
  larger than the estimated mass by \cite{Cova2019} even when the
  uncertainties in the radio power and in the correlation have been
  taken into account.

A strongly polarized signal ($\sim15-20\%$), associated with the radio
halo, has been detected by \cite{Girardi2016} across most of the radio
halo extension. Numerical three-dimensional simulations demonstrate
that an intracluster magnetic field with a strength $\langle B_0
\rangle \simeq 0.5\,\mu$G at the center of the cluster and fluctuating
over a range of scales between a few kpc up to 1\,Mpc is consistent
with both the total intensity and polarized emission of the radio halo
\citep{Girardi2016}.  The magnetic field has been constrained by the
fractional polarization of the diffuse emission, and is therefore relatively
independent of the relativistic electron population. Such a
magnetic field auto-correlation scale is much larger than the
resolution of radio observations ($\sim$130\,kpc), preventing the beam
depolarization of the radio signal that usually hinders the detection
of radio halo polarized emission \citep{Govoni2013}.  The value of the
magnetic field strength derived by \cite{Girardi2016} is consistent
with the results from upper limits on inverse Compton emission of the
cluster by \cite{Cova2019}, who derive a magnetic field larger than
0.2\,$\mu$G over the whole radio halo region and larger than
0.8\,$\mu$G when only the brightest region is considered.

\section{Radio observations and data reduction}
\label{observations}
For our analysis, we use new LOFAR data at 120-168\,MHz and new VLA observations in the frequency range 1-2\,GHz, as described in the following subsections.

\subsection{LOFAR observations}
We present LOFAR data (SAS ID 682162,672474) obtained through the observing program LC10\_024 (PI V. Vacca). A summary of the observations is reported in Table\,\ref{tab:B}. The observations have been carried out with the instrument in interferometer mode in the HBA Dual Inner configuration, with the full (core, remote and international stations) LOFAR array. The study presented in this paper is based on the data from the core and remote stations only. We used a 48\,MHz (120-168\,MHz) bandwidth with a total of 231 sub-bands per beam. The observations exploit the multi-beam capability of the HBA and were conducted together with the LoTSS observations through the co-observing program offered by the observatory. The observations presented here correspond to the LoTSS pointing P074+09. Due to the low declination of the source ($\sim 8.8^{\circ}$), the observation has been split in two 4\,h-blocks to ensure the elevation of the target is higher than 30$^{\circ}$ throughout the observation. 

\begin{table*}
\caption{Details of the radio brightness images presented in this work. Col.\,1, 2: central frequency and bandwidth; Col.\,3: uv-range; Col.\,4: resolution; Col.\,5: maximum angular scale accessible with the observations; Col.\,6: sensitivity; Col.\,7: robust; Col.\,8: figure/table where the image has been used.}
        \begin{tabular}{cccccccc}
         \hline
          \hline
Frequency & Bandwidth & uv-range&  Beam      & Max scale & $\sigma$&Robust& Figure/Table\\
MHz               & MHz       &$\lambda$ & ${\prime\prime}\times{\prime\prime}$& ${\circ}$      & mJy/beam&&\\
\hline
144               &48        &all        &  9$\times$9          &  1.2                    & 0.35  &-0.5& Fig.\,\ref{hr}\\
144               &48        &980- 16980 &  13.5$\times$13.5    &  0.06                   & 0.5   &-1& Table\,\ref{spix_ps}\\
144               &48        &all        &  20$\times$20        &  1.2                    & 0.4   &-0.5& Fig.\,\ref{a523}\\
144               &48        &158-16980  &  20$\times$20        &  0.36                   & 0.45  &-0.5& Fig.\,\ref{A523_spix} \\
144               &48        &all        &  65$\times$65        &  1.2                    & 1.3   &-0.5& Fig.\,\ref{a523vla}\\
144               &48        &158-4835   &  65$\times$65        &  0.36                   & 0.9   &-0.5& Fig.\,\ref{A523_spix}\\
144               &48        &196-4835   &  65$\times$65        &  0.29                   & 0.9   &-0.5& Fig.\,\ref{spix_plot}\\
1410              &192       &980- 16980 &  13.5$\times$13.5    &  0.06                   &0.09   &-1& Table\,\ref{spix_ps}\\
1410              &192       &158-16980  &  20$\times$20        &  0.36                   &0.075  &-0.5& Fig.\,\ref{A523_spix}\\
1410              &192       &all        &  65$\times$65        &  0.36                   &0.14   &0.5& Fig.\,\ref{a523vla}\\
1410              &192       &158-4835   &  65$\times$65        &  0.36                   &0.13   &-0.5& Fig.\,\ref{A523_spix}\\
1410              &192       &196-4835   &  65$\times$65        &  0.29                   &0.12   &-0.5& Fig.\,\ref{spix_plot}\\
1782              &320       &980- 16980 &  13.5$\times$13.5    &  0.06                   &0.08   &-1& Table\,\ref{spix_ps}\\
1782              &320       &all        &  65$\times$65        &  0.29                   &0.11   &0.5& Fig.\,\ref{a523vla}\\
1782              &320       &196-4835   &  65$\times$65        &  0.29                   &0.11   &-0.5& Fig.\,\ref{spix_plot}\\
\hline
\hline
        \end{tabular}
        \label{tab:images}

\end{table*}

To solve for the complex gains, and to calibrate the bandpass and
  amplitude, a direction-independent calibration step has been
  done. The non-directional calibration includes radio frequency
  interference (RFI) and bright off-axis source removal, as well as
  clock-total electron content separation based on the calibrator data
  \citep[see e.g.,][]{vanWeeren2016a}. During this step the
  calibrators J0813+4813 (3C196) and J0542+4951 (3C147) have been used
  as reference respectively for each of the two observing blocks.
  Direction-independent calibration of the data has been done with
  \emph{pre-factor} that takes 1\,s and 16\,ch/sb and averages down
to 8\,s and 2\,ch/sb \citep{vanWeeren2016a,deGasperin2019}. These
direction-independent calibrated datasets have then been processed
with the direction-dependent ddf-pipeline used by the LOFAR Surveys
Key Science Project (KSP) for reduction of the LoTSS survey
\citep{Tasse2020}\footnote{\url{https://github.com/mhardcastle/ddf-pipeline}}. The
ddf-pipeline is based on DDFacet \citep{Tasse2018} and KillMS
\citep{Tasse2014a,Tasse2014b,Smirnov2015} and repeats several
iterations of direction-dependent self-calibration towards 45
directions of the sky (i.e. facets), with an inner cut on the recorded
visibilities of 0.1\,km, corresponding to $\sim$48\,$\lambda$ at the
central frequency of the observations. Shorter baselines have
  been excluded because of RFI.  We performed a further refining of
the data in the region where the cluster is located through amplitude
and phase self-calibration. Flux contributions from sources outside a
square box with sides of length 0.4$^{\circ}$ centered at approximately the
  location of the X-ray peak of the cluster (RA 04h:59m:07s and Dec
  08$^{\circ}$:45$^{\prime}$:00$^{\prime\prime}$) have been subtracted in the \emph{uv}
plane using the direction-dependent calibrations and final
ddf-pipeline model, a phase-shift has been applied to locate the
target in the phase center, and finally a correction for the LOFAR
station beam towards this direction has been applied (this procedure
is detailed in \citealt{vanWeeren2020}).  We reimage the data that are
calibrated in the direction of our target using wsclean-2.10
\citep{Offringa2014}. A post-processing step, summarised in
\cite{Hardcastle2020} and which will be described in more detail in
Shimwell et al. in prep., was used to scale the wide-field image of
our pointing to align the fluxes with the same technique used for
LoTSS-DR2 which is itself aligned with the 6C survey
\citep{Hales1988,Hales1990} through a comparison with the \emph{NRAO
  VLA Sky Survey} \citep[NVSS, ][]{Condon1998}.  The flux scale of the
extracted data which, as described above, underwent further
processing, was then aligned with the wide field by applying an
overall multiplicative factor of 1.720 to the final images made from
the extracted data to ensure alignment consistent with the flux
scaling strategy used for the LoTSS Data Release 2.

\begin{figure*}
\includegraphics[width=0.9\textwidth]{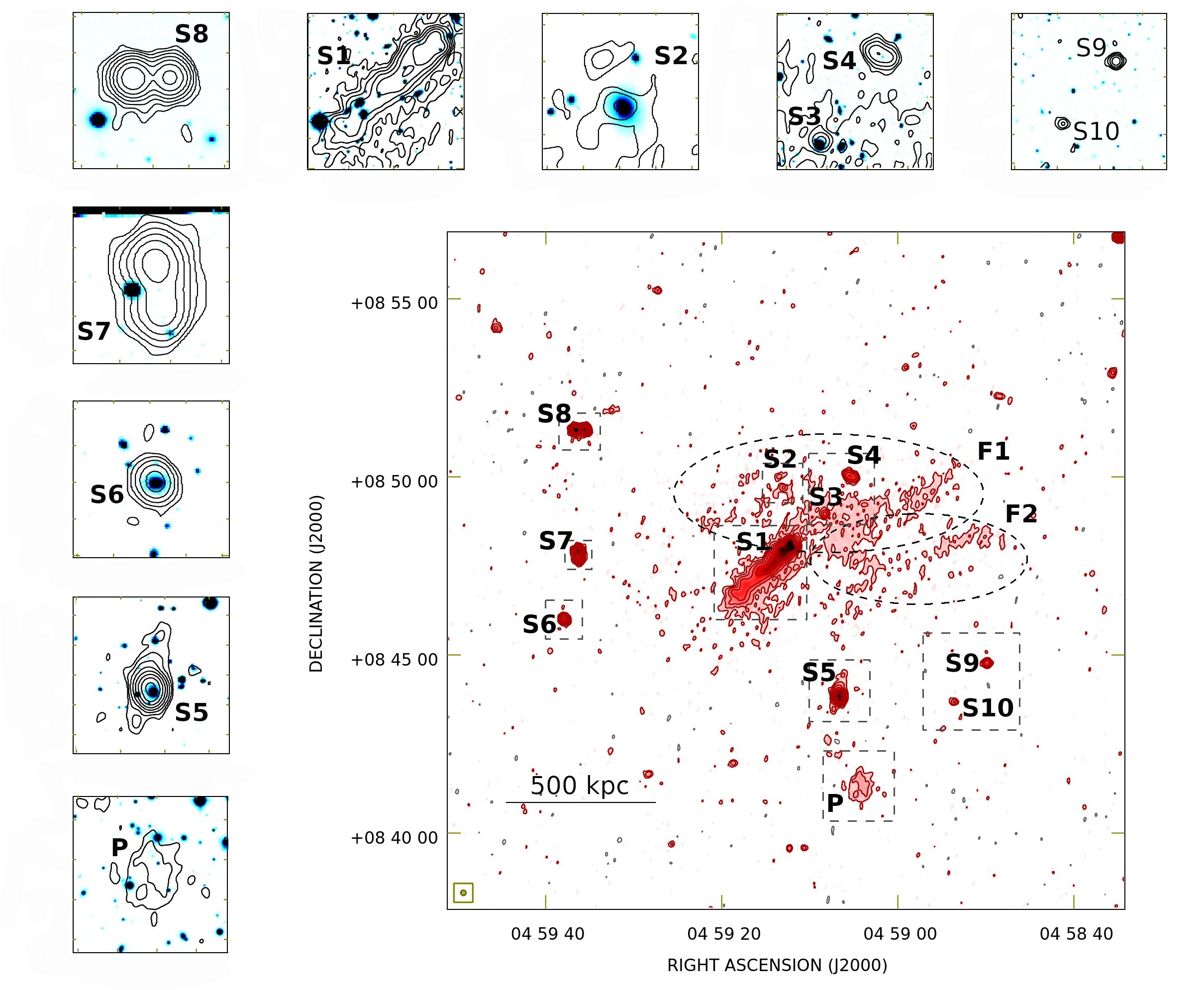}
\caption{\emph{Main panel:} LOFAR high-resolution image ($9^{\prime\prime}\times 9^{\prime\prime}$) of the central region of the galaxy cluster A523 in contours and colors. Red contours start at 3$\sigma$ ($\sigma$\,=\,0.35\,mJy/beam) and scale by a factor 2, gray contours are drawn at -3$\sigma$. The horizontal black bar corresponds to 500\,kpc at the chosen cosmology and the synthesised beam is shown in the bottom left. The dashed ellipses identify the position of the two filaments F1 and F2.  \emph{Left and top panels:} Zoom on the discrete sources in the field where the radio contours at 9$^{\prime\prime}$ are superimposed on the optical image from the INT r-band by \protect\cite{Girardi2016}, see the text for more details. Positive radio contours are the same as in the large panel. The position of the small panels is marked in the main panel by dashed gray boxes.}
\label{hr}
\end{figure*}

In order to investigate the emission from discrete sources as well as
from the radio halo, we produced images at low- and high-spatial
  resolution. At this declination the beam is not circular and, in
order to better compare our results with 1.4\,GHz observations we
convolved the high resolution image with a circular beam of
9$^{\prime\prime}$ and 13.5$^{\prime\prime}$, while circular beams of
20$^{\prime\prime}$ and 65$^{\prime\prime}$ have been used to restore
the images at lower resolution. We note that these images have been
produced including all baselines starting from
48\,$\lambda$. Shorter baselines have been excluded because of
  RFI, as
  noted above. Details about the images presented in the following are
  given in Table\,\ref{tab:images}. In the following, we assume an
uncertainty on the LOFAR flux density scale of 20\%, as done for
LoTSS \citep{Shimwell2019}.

\subsection{VLA observations}
\label{VLAobs}
We present new VLA data in the frequency range 1-2\,GHz in C and
D configuration obtained with a mosaic of six pointings in the
context of the observing program 13A-168 (PI M. Murgia). The details
of the observations are reported in Table\,\ref{tab:B}\footnote{In the
  table we report the coordinates of the central pointing only.}. The
data were reduced following standard procedures using the NRAO's
Astronomical Image Processing System (AIPS) package.  The data were
collected in spectral line mode in full Stokes, with a total
initial bandwidth of 1\,GHz subdivided in 16 spectral windows. 
  Each spectral window is 64\,MHz wide with 64 channels, and Hanning
smoothing was applied before bandpass calibration.  The source
J0319+4130 (3C84) was used as a bandpass and polarization leakage
calibrator, the source J0137+3309 (3C48) was used as a flux density
calibrator, and the nearby source J0459+0229 was observed for
  complex gain calibration. The flux density scale by \cite{Perley2017} was
adopted.  RFI flagging has been applied by excision of data with
values clearly exceeding the source flux ($\gtrsim 600$\,mJy).
Following calibration, the data were spectrally averaged to 16
channels of 4\,MHz per 64\,MHz wide spectral window. Spectral windows
from 0 to 5, 9 and 10 were too severely affected by RFI to provide
useful information, so we disregard them in the following.

Surface brightness images were produced using the Common Astronomy Software Applications (CASA) package (task tclean). The final images have been produced using spectral windows from 6 to 8 and from 11 to 15, including the full \emph{uv}-range available, unless differently stated. Circular restoring beams of 13.5$^{\prime\prime}$, 20$^{\prime\prime}$ and 65$^{\prime\prime}$ were used to restore the images in C, C+D, and in D configurations, respectively. Details about the images presented in the following are given in Table\,\ref{tab:images}. In the following, we assume an uncertainty on the VLA flux density scale of 2.5\%, in agreement with \cite{Perley2013}.

Everywhere in the text, unless differently stated, the overall uncertainties on the radio brightness and flux density include both the statistical uncertainty and the appropriate uncertainty in the flux density scale.

\section{Continuum images}
\label{lowfreq}

In Fig.\,\ref{hr} we show the 144\,MHz cluster emission with
9$^{\prime\prime}$ resolution. A number of radio galaxies, labelled
S1-S10, are present in the field embedded in the diffuse emission
  and with peaks in the radio brightness higher than 8\,$\sigma$. 
  Zoom in's on these radio galaxies are shown in the left and top
panels, where the radio contours are superimposed on the optical image
from the Isaac Newton Telescope (INT) r-band by
\cite{Girardi2016}. These authors identify the head tail S1 with
  the member-galaxy ID\,68, S3 with a point-like source with unknown
  redshift, S4 with a likely background galaxy, and S5 with the second
  BCG of the cluster. Using the optical information from the same
  authors, S2 appears to be the radio counterpart of the brightest
  cluster galaxy, previously undetected in radio, and S6 the dominant
  galaxy in the galaxy population of the ESE region, likely in the background at z$\sim$0.14.

\begin{figure*}
\includegraphics[width=0.8\textwidth]{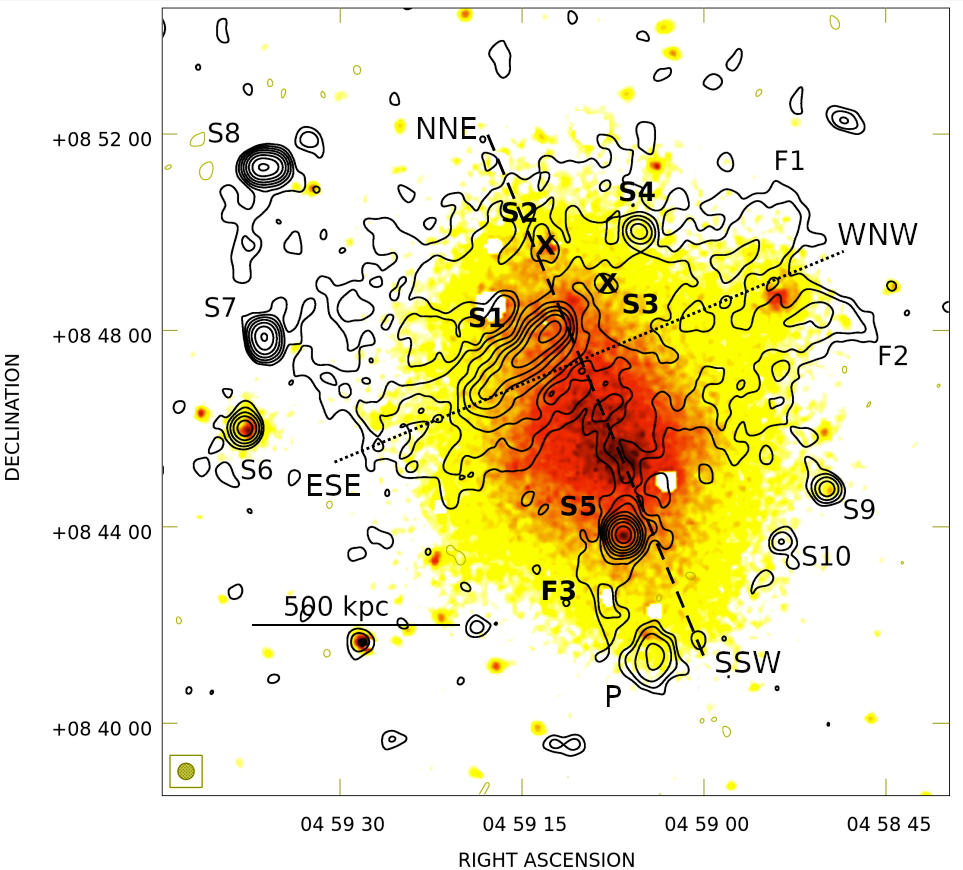}
\caption{Radio emission from the galaxy cluster A523 at 144\,MHz with LOFAR at 20$^{\prime\prime}$. Black contours start at 3$\sigma$ ($\sigma=$0.4\,mJy/beam) and scale by a factor 2, while negative 3$\sigma$ contours are drawn in army green. The beam is shown in the bottom left of the image and the horizontal black bar corresponds to 500\,kpc at the chosen cosmology. The radio emission is overlaid on the X-ray image of the cluster from XMM-Newton \protect\citep{Cova2019}, see the text for more details. The dashed and the dotted lines show respectively the axis of the primary and of the secondary merger.}
\label{a523}
\end{figure*} 

Two filaments of size $\sim$\,6$^{\prime}\,\times\,1^{\prime}$
(corresponding to about 710\,kpc\,$\times$\,120\,kpc) and with higher
surface brightness than their surrounding regions of radio emission
can be identified and are labelled respectively F1 and F2 (see dashed
ellipses in Fig.\,\ref{hr}). F1 occupies the region to the north-west
of S1, with a few embedded radio sources (S2, S3, S4). F2 is located
west of the central AGN S1. 
No other similar structures are seen elsewhere in the image,
  suggesting that these filaments are unlikely to be artefacts related
  to calibration or imaging but rather are real structures. Further
south, a new roundish patch of emission, labelled P, is detected at RA
04h59m04s and Dec 08$^{\circ}$41$^{\prime}$21$^{\prime\prime}$, with
an average brightness that is higher than the brightest regions of the
filaments.

\begin{table*}

\caption{Spectral index of point sources embedded in the diffuse emission. Col.\,1: Source name (see Fig.\,\ref{hr}); Col.\,2,3: RA and Dec corresponding to the peak of the source at 144\,MHz; Col.\,4, 5, 6: Flux density at 144\,MHz, 1.410\,GHz, and 1.782\,GHz respectively (this work); Col.\,7: Spectral index derived from the flux density values at 144\,MHz and 1.410\,GHz (this work); Col.\,8, 9, 10: Flux density from TGSS (147\,MHz) and NVSS (1.4\,GHz) and the corresponding spectral index as given by \protect\cite{deGasperin2018}.}
\scalebox{0.85}{
\begin{tabular}{ccccccccccc}

       \hline
        \hline
Source & RA (J2000) & Dec (J2000)&$S_{\rm 144\,MHz}$& $S_{\rm 1.410\,GHz}$& $S_{\rm 1.782\,GHz}$&$\alpha$& $S_{\rm TGSS}$& $S_{\rm NVSS}$&$\alpha_{\rm dG18}$\\
&hh:mm:ss&$^{\circ}$:$^{\prime}$:$^{\prime\prime}$&mJy&mJy&mJy&&mJy&mJy&\\
\hline\hline
S1& 04:59:12.3 & +08:48:04& 971.98$^1$ $\pm$ 194.43&   104.19$^1$  $\pm$ 2.68  &  89.25    $\pm$ 2.30   & 0.98 $\pm$ 0.09& 688.03$\pm$ 13.90 &121.71$\pm$1.22 & 0.77 $\pm$ 0.01\\
S2& 04:59:13.0 & +08:49:41 & 5.81    $\pm$ 1.17  &   0.54 $\pm$ 0.03& 0.23 $\pm$ 0.02   & 1.04 $\pm$ 0.09  & -&-& -\\
S3& 04:59:08.3 & +08:48:58 & 10.88   $\pm$ 2.18  &   1.04 $\pm$ 0.04&  0.93  $\pm$ 0.03 & 1.03 $\pm$ 0.09& -&- &-\\
S4& 04:59:05.4 & +08:50:01 & 32.48   $\pm$ 6.50  &   4.44  $\pm$ 0.11 &  4.55   $\pm$ 0.12  & 0.87 $\pm$ 0.09 &-&- &-\\
S5& 04:59:06.6 & +08:43:50& 313.68  $\pm$ 62.74 &   61.74   $\pm$ 1.54 &  55.84  $\pm$ 1.40  &0.71 $\pm$ 0.09 &213.91 $\pm$11.10& 61.33$\pm$ 1.01  &0.55 $\pm$ 0.02\\
S6& 04:59:37.9 & +08:46:01& 34.65   $\pm$ 6.93  &   6.91  $\pm$ 0.17 &  5.78   $\pm$ 0.15  &0.71 $\pm$ 0.09 & 0.0$\pm$ 0.0 & 6.11$\pm$ 1.02&0.52 $\pm$ 0.00 \\
S7& 04:59:36.3 & +08:47:54&112.34    $\pm$ 22.47 &   10.87 $\pm$ 0.27 &  8.20   $\pm$ 0.21  &1.02 $\pm$ 0.09 & 77.54$\pm$11.05& 8.69$\pm$ 1.02&0.97 $\pm$ 0.09 \\
S8& 04:59:36.5 & +08:51:20&238.71    $\pm$ 47.74 &   22.14 $\pm$ 0.57 &  17.35  $\pm$ 0.44 &1.04 $\pm$ 0.09   &144.93$\pm$ 11.41& 19.54$\pm$ 1.00 &0.89 $\pm$ 0.04\\
S9& 04:58:49.9 & +08:44:46&16.27   $\pm$ 3.26  &   3.21 $\pm$ 0.09&  2.83  $\pm$ 0.07 &0.71 $\pm$ 0.09   & 0.0 $\pm$0.0& 5.30$\pm$ 1.00&0.68 $\pm$ 0.00 \\  
S10& 04:58:53.6 &+08:43:41&4.87   $\pm$ 0.98  &   0.44 $\pm$ 0.02&  0.48  $\pm$ 0.02 &1.05 $\pm$ 0.09  &   -&-&-\\
                  \hline
           \hline
\multicolumn{9}{l}{\small $^1$ These flux density values include the contribution of the diffuse emission that amounts to about 35\,mJy at 144\,MHz and to about 2\,mJy at 1.4\,GHz.}
\end{tabular}

}
       \label{spix_ps}
\end{table*}

\begin{figure*}
\includegraphics[angle=0,width=0.95\textwidth]{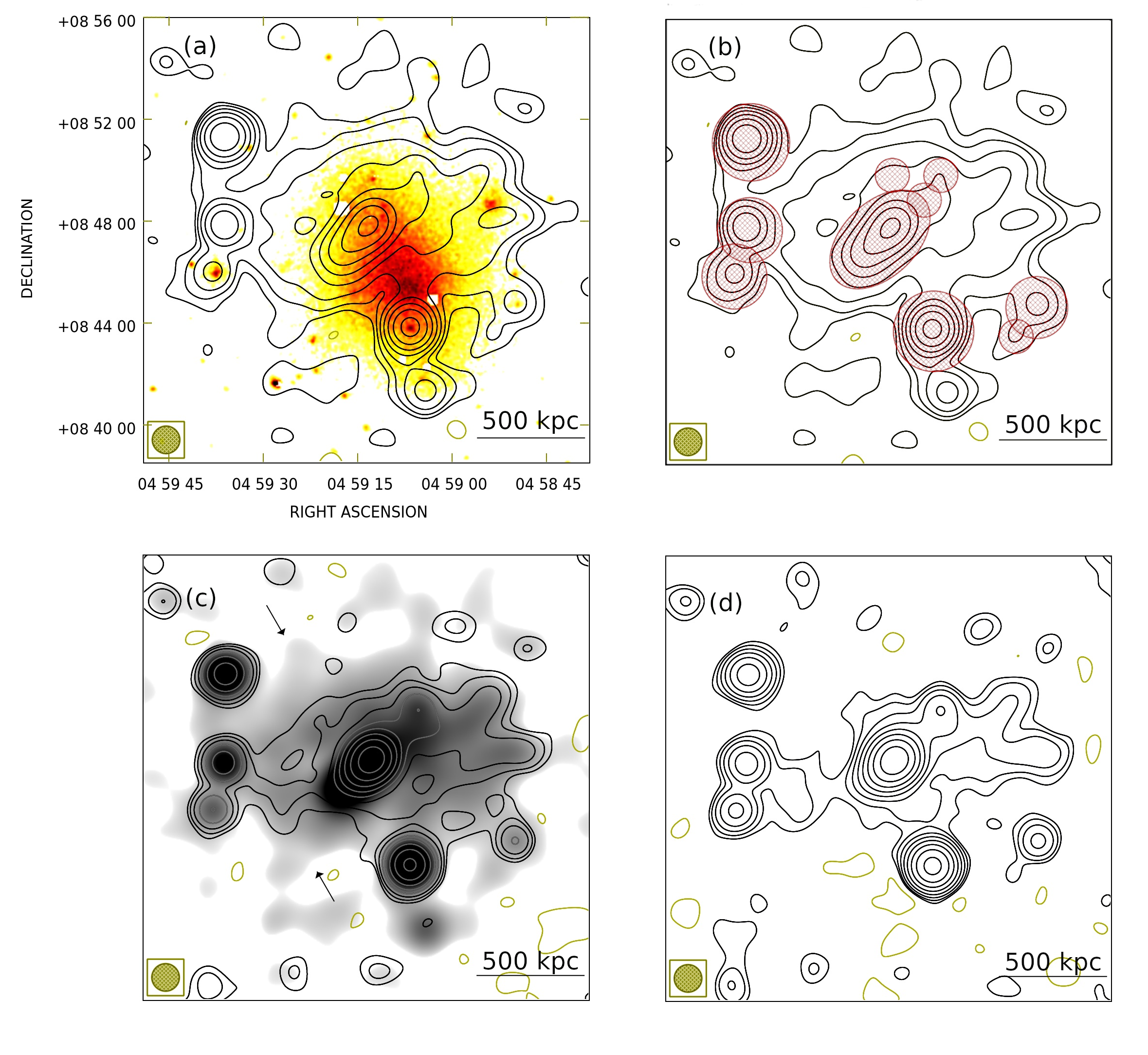}
\caption{
\emph{Panel (a)}: Radio emission in black contours from the galaxy cluster A523 at 144\,MHz ($\sigma=1.3$\,mJy/beam) overlaid on the X-ray image of the cluster from XMM-Newton \citep{Cova2019} in colors, see the text for more details. 
\emph{Panel (b)}: Contours are the same as in panel (a), with regions used to mask the emission of discrete sources shown in red.
\emph{Panel (c)}: Radio emission in black contours from the galaxy cluster A523 at 1.410\,GHz  ($\sigma=0.14$\,mJy/beam) overlaid on the radio emission in gray colors at 144\,MHz. Arrows point to the regions in the north-east and south-east detected with LOFAR but not with the VLA.
\emph{Panel (d)}: Radio emission in black contours from the galaxy cluster A523 at 1.782\,GHz ($\sigma=0.11$\,mJy/beam). 
All radio images are shown at 65$^{\prime\prime}$ of resolution and the synthesised beam is shown in the bottom left. Black contours start at 3$\sigma$ and scale by a factor 2, while negative 3$\sigma$ contours are drawn in army green. In each panel the horizontal black bar corresponds to 500\,kpc at the chosen cosmology.}
\label{a523vla}
\end{figure*} 

In Fig.\,\ref{a523} we show the radio image of the cluster at 144\,MHz obtained with LOFAR at 20$^{\prime\prime}$. The radio image is superimposed on the background-subtracted and exposure-corrected XMM-Newton image, with point sources removed. The X-ray image has been obtained with the combination of data from the three EPIC instruments in the soft 0.5-2.5\,keV band for a total exposure  time  of  about  220\,ks and was recently published by \cite{Cova2019}. Dashed and dotted lines indicate respectively the primary (SSW-NNE) and secondary (ESE-WNW) merger axis. The diffuse radio source extends further than  filaments F1 and F2, and a third filament, F3, is visible, with a size of $7^{\prime}\,\times\,1^{\prime}$ (i.e., 830\,kpc\,$\times$\,120\,kpc), that embraces S5 and terminates near the roundish patch of emission P. Overall, the diffuse emission sits in the northern part of the thermal gas distribution, is characterized by an E-W linear morphology that is approximately perpendicular to the axis of the main merger in the SSW-NNE direction, cuts through the central part of the X-ray region, and encompasses the discrete sources S1, S2, S3, S4, and S5 (see Fig.\,\ref{a523}). 

\subsubsection*{Flux density of point sources} 
In
  Table\,\ref{spix_ps}, we report the position and flux densities of
  the discrete sources S1-S10 at 144\,MHz, 1.410\,GHz, and 1.782\,GHz
  estimated from the images at 13.5$^{\prime\prime}$, obtained by
  applying the same \emph{uv}-range and weighting-scheme (robust=-1)
  at the three frequencies.  As in \cite{Vacca2018}, we model
  the sources with a 2-dimensional elliptical Gaussian sitting on a
  plane. Overall, there are nine free parameters of the fit: the x, y
  coordinates in the sky of the centre of the Gaussian, the full width
  half maximum along the two axes, the position angle, the amplitude,
  and the three components of the direction normal to the plane.  The
  non-zero baseline fit ensures that any contribution from the diffuse
  emission, if present, is not absorbed into the flux density of the
  sources. The uncertainty has been estimated by adding in
  quadrature the uncertainty derived from the fit procedure and the
  flux scale uncertainty at the corresponding frequency (20\% at
  144\,MHz and of 2.5\% at 1.410 and 1.782\,GHz). This procedure was
  used for all the discrete sources except the central AGN S1. Being
  an extended source, we determined the flux density by applying a
  3$\sigma$ sensitivity threshold to the LOFAR image
  ($\sigma=0.5$\,mJy/beam) and blanked the two  VLA images in the
  same regions as the LOFAR one. We refer to \S\,\ref{spixds} for
  spectral studies of these sources.

\begin{figure}
\includegraphics[angle=0,width=0.5\textwidth]{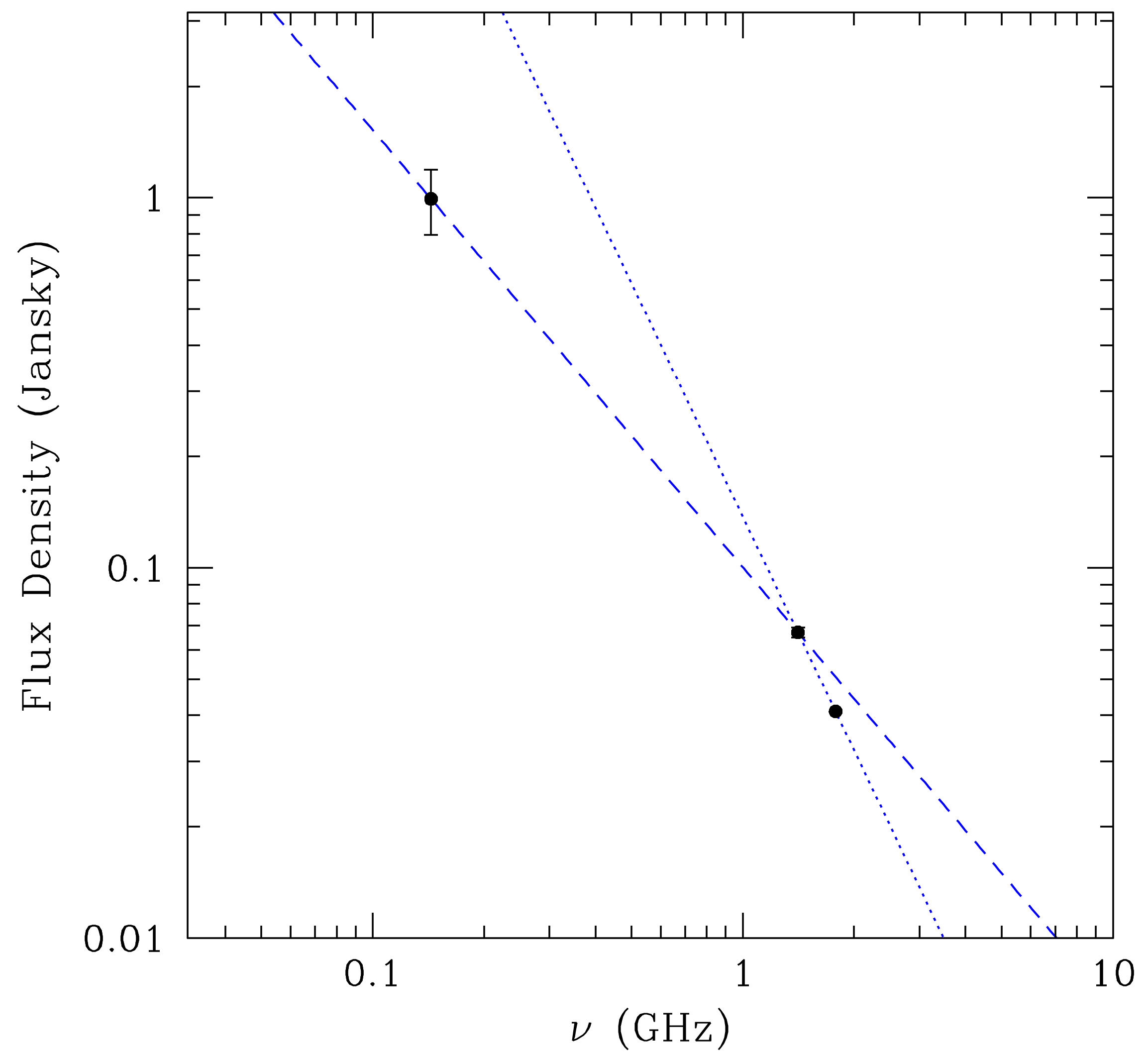}
\caption{Flux density of the diffuse emission in A523 versus frequency. The power-law fit of the 144\,MHz and 1.410\,GHz data points is shown in dashed blue, while the fit of the 1.410\,GHz and 1.782\,GHz data points in dotted blue.}
\label{spix_plot}
\end{figure}
\begin{table*}
\caption{Details of the spectral index maps. Col.\,1, 2: minimum and maximum cut in \emph{uv}-range; Col.\,3, 4: minimum and maximum angular scales accessible with the data; Col.\,5: original resolution of the 144\,MHz image; Col.\,6: original resolution of the 1.410\,GHz image; Col.\,7: restored beam.}
        \begin{tabular}{ccccccc}
         \hline
          \hline
\emph{uv}-min& \emph{uv}-max& $\theta_{\rm min}$ &$\theta_{\rm max}$ & Beam$_{\rm 144\,MHz}$ &Beam$_{\rm 1.410\,GHz}$& Beam$_{\rm restored}$\\
$\lambda$& $\lambda$& $\prime\prime$ &$\prime$ & ${\prime\prime}\times{\prime\prime}$& ${\prime\prime}\times{\prime\prime}$& ${\prime\prime}\times{\prime\prime}$\\
\hline
158 & 4835  & 48 & 22  &$34.5\times46.6$&$36.1\times49.8$&$ 65\times65$\\
158 &16980  & 14& 22 &$7.8\times15.9$&$13.0\times16.3$&$20\times20$\\
\hline
\hline
        \end{tabular}
        \label{tab:spix}

\end{table*}
\subsubsection*{Flux density of diffuse emission}
In panel (a) of Fig.\,\ref{a523vla} we show in black contours the LOFAR image of the cluster at 144\,MHz overlaid on the same X-ray image as in Fig.\,\ref{a523}, while in panel (b) the LOFAR image with the regions used to mask discrete sources. In the panels (c) and (d), we show on the left the new VLA image at 1.410\,GHz in contours overlaid on the LOFAR image in gray scale and on the right the new VLA image at 1.782\,GHz. 
All the radio images are at 65$^{\prime\prime}$ resolution.
Comparison of the LOFAR and VLA images shows that the radio halo appears more extended at lower frequencies. The LOFAR image at 65$^{\prime\prime}$ indeed reveals a size of the source of about 15$^{\prime}$, corresponding to 1.8\,Mpc at the cluster distance. The maximum angular scale accessible with our data is about 1.2$^{\circ}$, 22$^{\prime}$, and 17$^{\prime}$ respectively at 144\,MHz, 1.410\,GHz and 1.782\,GHz, as reported in Table\,\ref{tab:images}. The sensitivities $\sigma$ at these three frequencies and spatial resolution are 1.3\,mJy/beam, 0.14\,mJy/beam, and 0.11\,mJy/beam.  A 
brightness of 3.9\,mJy/beam at 144\,MHz (i.e., at 3$\sigma$), as the faintest regions of the emission in the LOFAR image, corresponds to 0.400-0.160\,mJy/beam at 1.410\,GHz and to 0.315-0.115\,mJy/beam at 1.782\,GHz, for typical spectral index values ($\alpha=1-1.4$). A sensitivity $\sigma$ of 0.133-0.053\,mJy/beam at 1.410\,GHz and of 0.105-0.038\,mJy/beam at 1.782\,GHz is needed to detect this emission, while the sensitivity in our images is 0.14\,mJy/beam and 0.11\,mJy/beam respectively at 1.410\,GHz and 1.782\,GHz, see Table\,\ref{tab:images}. The larger extent detected at LOFAR frequencies compared to VLA higher frequency images is therefore likely due to the better sensitivity to steep spectrum emission of LOFAR.

\begin{figure*}
\includegraphics[width=1.0\textwidth]{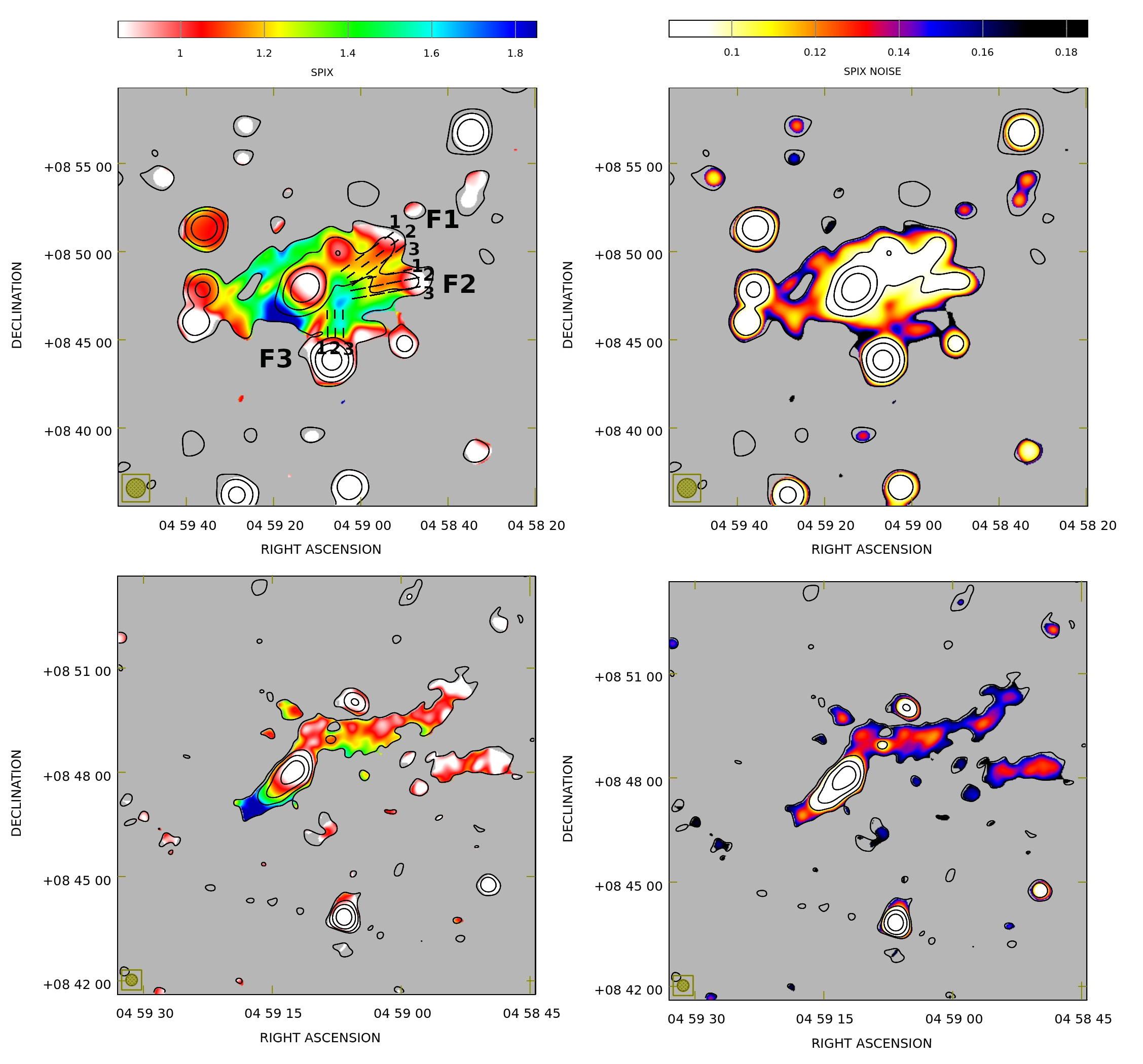}
\caption{\emph{Top panels}: Spectral index (left) and spectral index uncertainty (right) image of A523 between 144\,MHz and 1.410\,GHz at 65$^{\prime\prime}$ resolution. Contours represent the radio emission at 1.410\,GHz at the same spatial resolution, start at 3$\sigma$ and increase by a factor 4 ($\sigma=0.13$\,mJy/beam). 
\emph{Bottom panels}: Spectral index (left) and spectral index uncertainty (right) image of the central region of A523 between 144\,MHz and 1.410\,GHz at 20$^{\prime\prime}$ resolution. Contours represent the radio emission at 1.410\,GHz at the same spatial resolution, start at 3$\sigma$ and increase by a factor 4 ($\sigma=75\,\mu$Jy/beam).}
\label{A523_spix}
\end{figure*}

A faint region of diffuse emission in the north-east and in the south
of the cluster emerges in the 144\,MHz image at 65$^{\prime\prime}$,
along the main merger axis (see arrows in Fig.\,\ref{a523vla}),
  panel (c), along the optical and X-ray distribution of the system, see Fig.\,1 in \cite{Girardi2016}. Consequently, at this frequency the emission along
  the SSW-NNE axis is more prominent than at $\sim$1-2\,GHz.  At
1.410\,GHz only the brightest regions along the ESE-WNW direction
survive, while the emission in the south, in the south-east and in the
north-east falls below the noise.  At 1.782\,GHz, the diffuse emission
dims further, the filaments F1, F2, part of F3 and a bridge of
radio emission connecting S1 to S6 only are still apparent. In
order to measure the flux densities at the three frequencies, we
masked embedded compact sources as shown in Fig.\,\ref{a523vla} (panel
 b). We assume that the halo extends over the location of
embedded compact sources. In these masked regions we assume the halo
radio brightness is equal to its mean radio brightness over the entire
unmasked diffuse source. Overall, we derive a flux density at 144\,MHz
of $S_{\rm 144\,MHz}\,=\,(1.52\,\pm\,0.31)$\,Jy, including the
southern patch. The uncertainty takes into account the
  uncertainty on the flux scale and the image noise. By using the
same method to extrapolate over embedded masked regions, at 1.410\,GHz
we measure a flux density of the diffuse emission of $S_{\rm
  1.410\,GHz}\,=\,(70\,\pm\,2)$\,mJy, in agreement within
  $\approx$\,0.8$\sigma$ with the flux
density derived by \cite{Girardi2016} $S_{\rm
  1.4\,GHz}\,=\,(72\,\pm\,3)$\,mJy, while at 1.782\,GHz we measure a flux
density $S_{\rm 1.782\,GHz}\,=\,(42\,\pm\,1)$\,mJy.  These flux
densities have been derived considering all the available
\emph{uv}-range and applying a sensitivity cut of 3$\sigma$ at the
corresponding frequency, so they refer to the full size of the source
at those frequencies.

In panel (a) of Fig.\,\ref{a523vla}, the patch P is more
extended compared to Fig.\,\ref{a523} and blended with S5 and
  with the rest of the diffuse emission. In order to better
  discriminate the emission associated with this source, we used
  the 20$^{\prime\prime}$ image. We measure a flux density $S_{\rm
    144\,MHz}\,=\,(49\,\pm\,10)$\,mJy and an average brightness
  $4.5\,$mJy/beam. While this patch is clearly detected at
  144\,MHz, only an excess is present at the $3\sigma$ level
  ($\approx\,0.4$\,mJy/beam) at 1.410\,GHz, suggesting a spectral
  index $\alpha\gtrsim1.5$. The origin of this emission will be
  discussed in \S\ref{Spectral_index_images}.

\section{Spectral index}
\label{spix_sect}

In Fig.\,\ref{spix_plot} we plot the flux density of the diffuse emission at 144\,MHz, 1.410\,GHz, and 1.782\,GHz versus frequency, derived from images obtained by selecting the same \emph{uv}-range (196-4835$\lambda$), weighting-scheme (robust=-0.5), and area on the sky. The $uv_{\rm min}=196\lambda$ is imposed by the minimum uv-sampling at 1.782\,GHz. The corresponding images at 144\,MHz, 1.410\,GHz, and 1.782\,GHz are not shown here.
Overall, the spectrum cannot be described with a single power-law, since it steepens towards high frequency. Indeed, as discussed in \S\,\ref{lowfreq}, at 1.782\,GHz only the filaments F1, F2, part of F3, and a bridge connecting S1 to S6 survive, the rest of the emission is buried by the noise. We fit the spectrum  with the software \emph{synage} \citep{Murgia1996} assuming a power-law.
Between 144\,MHz and 1.410\,GHz we obtain $\alpha_{\rm 144\,MHz-1.410\,GHz}=1.2_{-0.2}^{+0.1}$ (see dashed line in Fig.\,\ref{spix_plot}), while between 1.410\,GHz and 1.782\,GHz we obtain $\alpha_{\rm 1.410\,GHz-1.782\,GHz}=2.1_{-0.3}^{+0.2}$  (see dotted line in Fig.\,\ref{spix_plot}). A similar behaviour has been observed for other radio halos \citep[e.g.,][]{Deiss1997,Thierbach2003,Xie2020,Rajpurohit2021}. However, our frequency range is smaller than in these studies, where it extends up to about 5\,GHz.

In order to perform a detailed analysis of the spectral index
properties of the diffuse emission, we produced spectral index images
at low- and high-spatial resolution with our new LOFAR observations at
144\,MHz and the VLA data at 1.410\,GHz presented in
\S\,\ref{observations}.  We selected the same \emph{uv}-range and
weighting-scheme (robust=-0.5) at the two frequencies both at high and
at low spatial resolution and a beam of $65^{\prime\prime}$ and
20$^{\prime\prime}$ was restored, respectively, in the two cases. All
the details about the selected \emph{uv}-range, the corresponding
angular scales accessible through the data, the original spatial
resolution of the images and the size of restored beams are reported
in Table\,\ref{tab:spix}. All the images have been regridded to a
  common frame in order to have the same pixel size.

The spectral index maps have been obtained by applying a sensitivity cut in radio brightness of 3$\sigma$ at both frequencies simultaneously, in order to exclude 
pixels that are below 3$\sigma$ in at least one of the two images. Flat-spectrum regions above the noise level at 1.410\,GHz but not at 144\,MHz and steep-spectrum regions detected at 144\,MHz but not at 1.410\,GHz will be missed with this procedure. In particular, the peripheral regions that are detected at 144\,MHz but not at 1.410\,GHz are characterized by a spectral index steeper than $\alpha\gtrsim 0.9$. In order to investigate the spectral index behaviour also in these regions, we directly compare the brightness profile at 144\,MHz and 1.410\,GHz without applying any sensitivity cut to the images.

\subsection{Spectral index images}
\label{Spectral_index_images}

The spectral index distribution (left) and its uncertainty (right) are
shown in Fig.\,\ref{A523_spix} at both low (65$^{\prime\prime}$, top
panels) and high (20$^{\prime\prime}$, bottom panels) spatial
resolution. The spectral index trends observed at low resolution are
consistent with those observed at higher spatial resolution. However,
at 20$^{\prime\prime}$ resolution, only the brightest parts of the
emission are visible.

The low-resolution spectral index image between 144\,MHz and
  1.410\,GHz reveals a patchy distribution with a flattening in the
  west at the location of the filaments, and a steepening both in the
  north-east and in the south-east. 
  In particular, the region in
the south-east of S1 is characterized by a spectral index $\alpha\sim
1.2-2.2$, likely because it is contaminated at least in part by the
steep spectrum tails of S1. The north-east and south of the diffuse
emission appear well detected in the 144\,MHz image but fall below the
noise at 1.410\,GHz, with the brightest patches at 144\,MHz reaching
radio brightness up to 20\,mJy/beam. If we take a brightness
  upper limit of 0.39\,mJy/beam at 1.410\,GHz (i.e., 3$\sigma$ with
$\sigma=0.13$\,mJy/beam), we derive a spectral index  $\alpha >
  1.7$.

\begin{figure}
\includegraphics[width=0.4\textwidth]{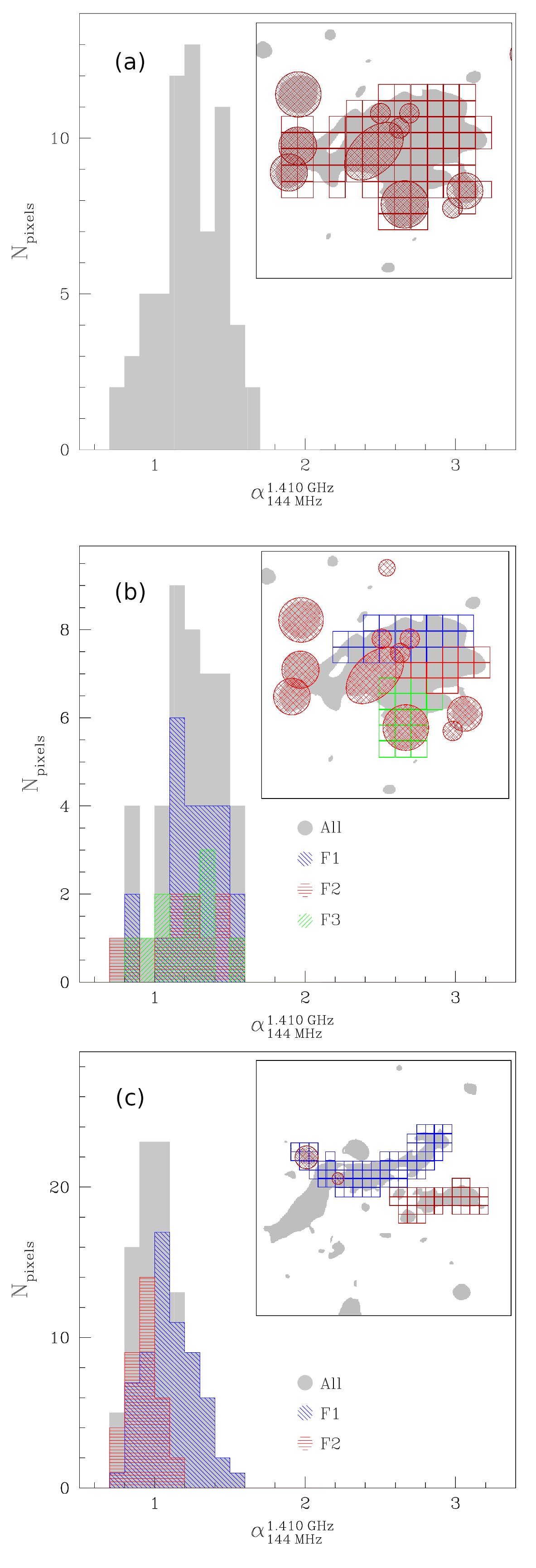}
\caption{\emph{Panel (a)}: Histogram of the low-resolution spectral index image obtained covering the image with the grid shown in the inset and masking discrete sources. \emph{Panel (b)}: Histogram of the spectral index values at the location of the filaments at 65$^{\prime\prime}$ (overall in gray, F1 only in blue, F2 only in red, and F3 only in green). \emph{Panel (c)}: Histogram of the spectral index values at the location of the filaments at 20$^{\prime\prime}$ (overall in gray, F1 only in blue, and F2 only in red). In all cases the side of the cells of the grid has a size equal to the beam FWHM. }
\label{A523_histo_spix}
\end{figure} 

\begin{table}
\center
\caption{Spectral index values for the full diffuse emission, the filaments and along each filament. Col.\,1: Region considered for the average; Col.\,2, 3, 4: average spectral index $\langle\alpha\rangle$, its dispersion $\sigma_{\alpha}$ and the average uncertainty $\langle \sigma_{\alpha}\rangle$; Col.\,5: spatial resolution of the images used for the measurements.}
        \begin{tabular}{ccccc}
         \hline
          \hline
Source & $\langle\alpha\rangle$& $\sigma_{\alpha}$ & $\langle \sigma_{\alpha}\rangle$&Beam ($\prime\prime$)\\
\hline\hline
All&1.2&0.2&0.1& 65\\
Filaments&1.2&0.2&0.1& 65\\
F1&1.3&0.2&0.1& 65\\
F2&1.2&0.3&0.1& 65\\
F3&1.2&0.2&0.1&65\\
Filaments&1.0&0.2&0.1& 20\\
F1&1.1&0.2&0.1& 20\\
F2&0.9&0.1&0.2& 20\\
              
                  \hline
            \hline
        \end{tabular}
        \label{spix_average}

\end{table}

\begin{figure}
\includegraphics[width=0.45\textwidth]{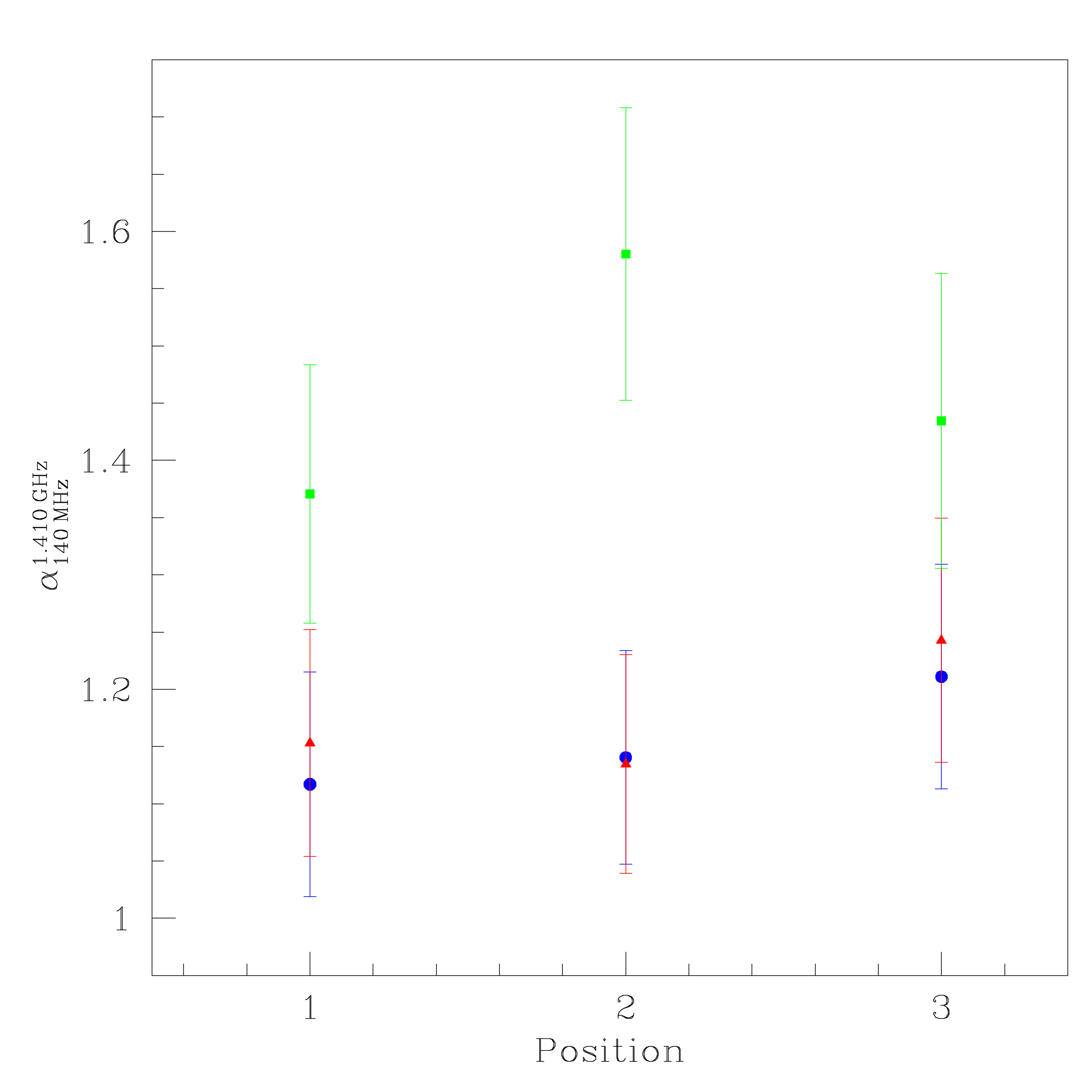}
\caption{Spectral index along filament F1 (blue dots), filament F2 (red triangles) and filament F3 (green squares). The slices are shown in the top left panel of Fig.\,\ref{A523_spix} and the numbers on the x-axis indicate the slice identification number as reported in the same figure, see the text for more details.}
\label{A523_slices_spix}
\end{figure} 

In Fig.\,\ref{A523_histo_spix} (panel (a)) we show the histogram of
the spectral index values of the diffuse emission obtained covering
the low resolution spectral index image with a grid, after
  masking compact sources, as shown in the inset. The side of the
  cells is equal to the full width at half maximum (FWHM) of the
  beam. We derive an average spectral index
$\langle\alpha\rangle=1.2$, a dispersion $\sigma_{\alpha}=0.2$, and an
average value of the spectral index statistical
uncertainty $\langle\sigma_{\alpha}\rangle=0.1$, see
Table\,\ref{spix_average}\footnote{Values derived by weighting the
  average radio brightness both at 144\,MHz and at 1.410\,GHz in the
  corresponding cell are in agreement with these values within the
  uncertainty.}. The average spectral index statistical
uncertainty has been derived from the spectral index uncertainty
image (top right panel in Fig.\,\ref{A523_spix}). This image has been
obtained by evaluating the statistical uncertainty on a pixel
basis, including both the flux density scale uncertainty and the
thermal noise of the two radio images. If the patchy structure of the
spectral index image is due to measurement uncertainties, we
expect that the mean value of the uncertainty image and the
dispersion of the spectral index distribution are consistent. The
dispersion is only slightly higher than the uncertainty derived
by using the spectral index noise image suggesting that, even if
  some degree of intrinsic complexity is present, the observed
fluctuations are dominated by the measurement uncertainties.
Following \cite{Orru2007}, since the total dispersion is the result of the sum in quadrature of the uncertainty and of the intrinsic scatter 
\begin{equation}
\sigma_{\alpha}=\sqrt{\langle\sigma_{\alpha}\rangle^2 +\sigma_{\rm int, proj}^2}, 
\end{equation}
we derive an intrinsic scatter in the spectral index $\sigma_{\rm int, proj}=0.17$ projected in the plane of the sky on scales of the beam, i.e. 65$^{\prime\prime}$. If the spectral index varies stochastically along the line of sight over the full diffuse emission region, we derive a 3-dimensional intrinsic spectral index $\sigma_{\rm int}=\sqrt{N}\sigma_{\rm int, proj}=0.67$, where $N=15$ is the number of cells assuming that the maximum size of the source along the line of sight is similar to the largest linear size in the plane of the sky, i.e., 15$^{\prime}$. This is comparable to the value derived by \cite{Botteon2020a} for the emission in A2255 and higher than the value derived for the radio halos in the Toothbrush cluster by \cite{vanWeeren2016} and in A520 by \cite{Hoang2019} by a factor of $\sim 7$ and 2 respectively.

\subsubsection*{Filaments}

In the west of the source, the spectral index distribution shows a flattening at the location of the three filaments. Histograms of the spectral index distribution in this region are shown in the panels (b) and (c) of Fig.\,\ref{A523_histo_spix}. 
In the central panel, we show the statistics derived from the low-resolution image, using the LOFAR image at 20$^{\prime\prime}$ to determine the location and size of the filaments. Considering all the filaments simultaneously, the average spectral index is $\langle\alpha\rangle=1.2\pm 0.1$. When considered separately, we derive respectively $\langle\alpha\rangle_{\rm F1}=1.3\pm 0.1$, $\langle\alpha\rangle_{\rm F2}=1.2\pm 0.1$, and $\langle\alpha\rangle_{\rm F3}=1.2\pm 0.1$, see Table\,\ref{spix_average}. However, the low resolution image hinders a detailed analysis of the spectral properties of the filaments, due to a blending with the rest of the diffuse emission.
In the panel (c) of Fig.\,\ref{A523_histo_spix}, we show the histograms of the spectral index distribution along the filaments obtained from the high-resolution image. We carried out our measurements only for F1, still clearly visible, and for the brightest region of F2. Only a small patch of F3 survives, therefore we did not include it in the analysis. 
We find an overall average spectral index $\langle\alpha\rangle=1.0\pm 0.1$, and   $\langle\alpha\rangle_{\rm F1}=1.1\pm 0.1$ and $\langle\alpha\rangle_{\rm F2}=0.9\pm 0.2$, respectively for F1 and F2, see Table\,\ref{spix_average}. 
Even if still in agreement within the measurement uncertainties, the spectral index in the brightest regions of these filaments appears slightly flatter than the rest of the source, and in particular, the emission along the filaments F2 is characterized by a flatter spectrum than that along F1. However, this difference could be due to the fact that in the high resolution image we are only probing the brightest regions of F2, while for F1 more structure is used.

A qualitative inspection of the 65$^{\prime\prime}$ image
  indicates that moving away from the filament brightest region, the
  spectral index becomes steeper. To quantify this behaviour, we
  computed the spectral index along slices in the low-resolution
  image. For each filament we used three slices: a central one located
  along the brightest part of the filament and two others, one on each
  side. Slices are one pixel in width, spaced by half FWHM (i.e.,
  32.5$^{\prime\prime}$) and with lengths comparable to the length of
  the filaments. To better identify the filaments we used the high
  resolution LOFAR image at 20$^{\prime\prime}$. In
  Fig.\,\ref{A523_slices_spix}, we show the average spectral index
  along the three slices for the three filaments: filament F1 in blue
  dots, filament F2 in red triangles and filament F3 in green
  boxes. The location of the slices is shown in Fig.\,\ref{A523_spix}
  and is identified with the numbers 1, 2, and 3 (numbered from North to South for F1 and F2 and from left to right for F3), as reported in the
  x-axis of Fig.\,\ref{A523_slices_spix}.  F1 shows a spectral
  index decreasing from south to north, while F2 shows a spectral index
  slightly steeper moving towards both north and south. However, within the errors,
  the two filaments show similar spectral indices with variations between 1 and
  1.3. Finally, F3 has a steeper spectral index that varies in the range
  1.3-1.7 with increasing values at the location of the brightest
  patches of the emission.
  The different spectral behaviour of F1 and F2 with respect to the surrounding medium could be related to a stronger magnetic field that allows us to detect lower-energy electrons with a flatter spectrum.
  Spectral index images of these filaments at
  higher resolution are necessary to investigate in more depth their
  behaviour, with no contamination from the rest of the diffuse
  emission.

Filaments of diffuse emission have also been observed in the
  galaxy cluster A2255. \cite{Govoni2005} classified them as
  peripheral structures associated with the radio halo in agreement
  with expectations from numerical simulations \citep{Loi2019}, while
  \cite{Pizzo2011} classified them as relics, due to their morphology,
  fractional polarization level and rotation measure. The filaments
  in A2255 show spectral indices flatter ($\alpha\sim 0.8-1.3$) than the
  rest of the diffuse emission (up to $\alpha\sim 2$), see also
  \cite{Botteon2020a}. Similar filamentary emission has been recently
  identified in the diffuse emission in MACS\,J0717.5+3745
  characterized by a spectral index of $\alpha\sim 1.2$
  \citep{Rajpurohit2021a}. \cite{Rajpurohit2021} speculate that such
  features may be due to the complex distributions of shocks caused by
  the merger in the cluster.
  In line with this, the filaments in A523 could be either radio
  relics or associated with the peripheral regions of the radio
  halo. Discriminating between the two is not a trivial task. F1 and
  F2 could be associated with shock waves produced by the main merger
  along the SSW-NNE direction, while F3 could be associated with a
  shock wave produced by the secondary merger along the perpendicular
  axis. Shock fronts have not been detected at the sensitivity of
  X-ray observations currently available and, only in case of the
  filament F1, the morphology and the spectral index distribution
  could support this possibility. In the case of a merger with some
  inclination with respect to the plane of the sky, the detection of
  shock waves and of a spectral gradient could be hindered by
  projection effects. The available X-ray and optical data do not show
  evidence of a merger component along the line of sight.  Such a
  component could be possibly associated with the secondary
  merger. However, at the moment, no indication of this emerges from
  the data. For a detailed description of the sub-clumps and of the
  merger geometry see \cite{Girardi2016} and \cite{Cova2019}.

\begin{figure*}
\includegraphics[width=0.7\textwidth]{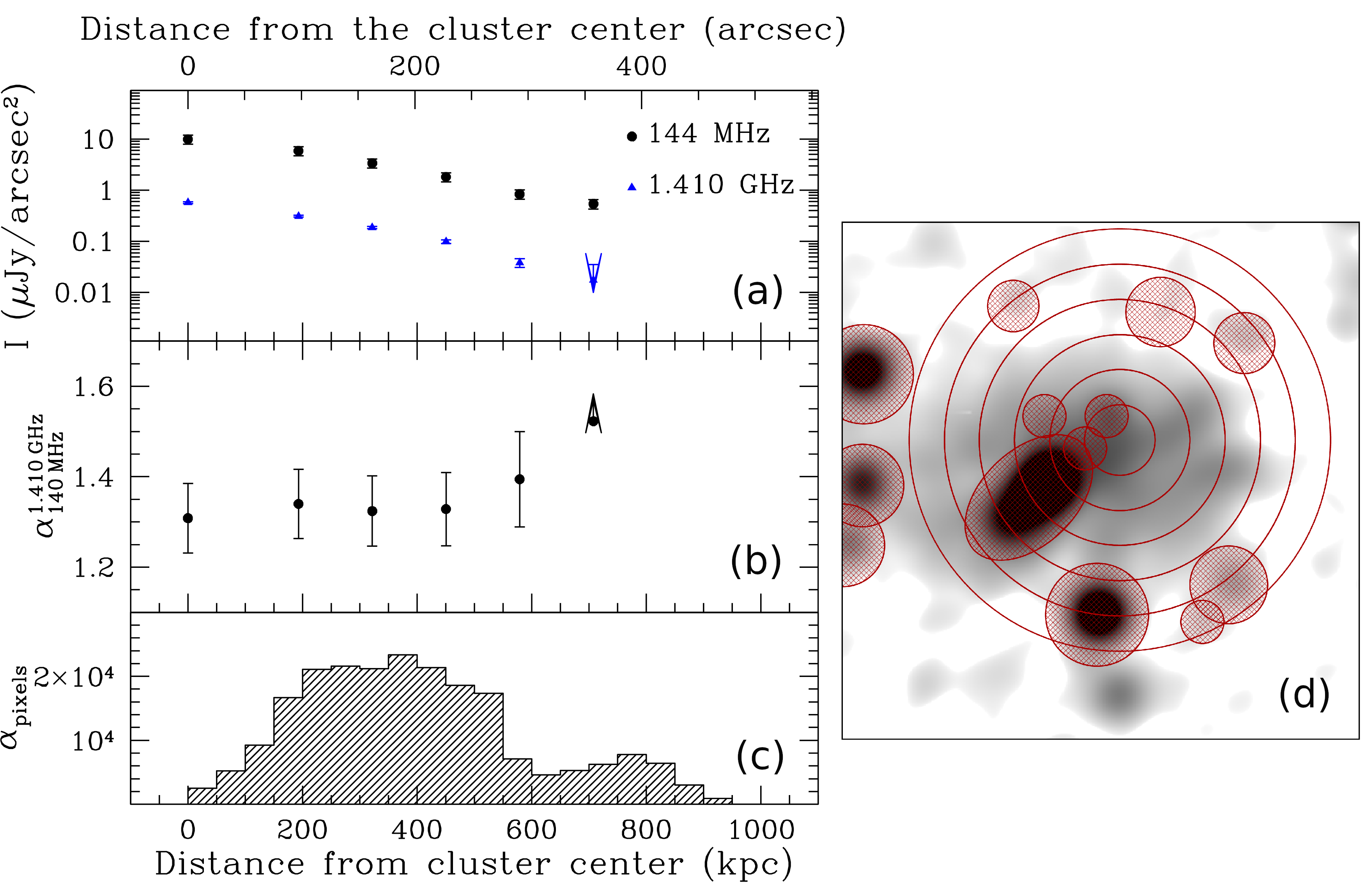}
\caption{\emph{Panel (a)}: Radial profile of the radio brightness at 144\,MHz (black dots) and 1.410\,GHz (blue triangles) with respect to the radio peak, using the annuli shown in the panel (d). \emph{Panel (b)}: Radial profile of the spectral index of the diffuse emission obtained from the radio brightness profiles in the panel (a). \emph{Panel (c)}: pixel distribution of the spectral index 65$^{\prime\prime}$ image as a function of the distance from the radio peak at 144\,MHz. The pixel distribution does not depend on the annuli chosen to produce the radial profiles shown in the above panels. \emph{Panel (d):} Gray scale shows the LOFAR image at 144\,MHz at 65$^{\prime\prime}$, as well as annuli with a beam-size width centered on the radio peak at 144\,MHz and used in order to derive the radial profile. Discrete sources have been masked and therefore not included in the measurements.}
\label{A523_spix_profile}
\end{figure*}

\subsubsection*{Patch P}

Patch P is clearly detected at 144\,MHz at about RA ${\rm 04h:59m:04s}$ and Dec ${\rm +08^{\circ}:41^{\prime}:23^{\prime\prime}}$, while only the peak of
the emission associated with it is visible at 1.410\,GHz, with a
brightness of $\approx$\,0.4\,mJy/beam. If we compare this value with
the peak radio brightness at 144\,MHz, i.e. $42\,$mJy/beam, we obtain
a spectral index $\alpha=2.0$ in the brightest region of the emission.
This steep value might suggest that the patch P could be a very
old object, relic emission from a radio lobe, or plasma emitted by a
now-quiescent AGN.  
  The closest radio galaxy is the source S5 \citep[z=0.1041,
  ][]{Girardi2016} which shows signs of possible re-starting jets when
  inspected at high spatial resolution (see Fig.\,\ref{hr}). Patch P
  is distant almost $\sim 2.5^{\prime}$ from S5, corresponding to
  a projected linear distance of about 300\,kpc. If patch P is a remnant lobe of
  previous activity of S5, it could have moved up to the distance
  where we observe it now. In this case, its linear size would be
  about 170\,kpc. Alternatively, if we select close optical
    sources from \cite{Girardi2016} and \cite{Golovich2019} within a
  radial distance of one beam (65$^{\prime\prime}$) from the peak of
  patch P, we find 18 sources with redshift 0.097$\lesssim z\lesssim
  $0.637, translating into a linear size of the source between
  170\,kpc and 640\,kpc\footnote{We measure a largest angular size of $\sim$1.5$^\prime$ in Fig.\,\ref{a523}}, the closest being a source at
  z$\approx$0.637. According to radiative cooling models, old remnant
  sources are expected to be characterized by very steep spectra
  \citep[$\alpha\gtrsim1.2$, ][]{Pacholczyk1970} and observations of
  the Lockman Hole field with LOFAR show that these sources
  represent a few percent of the low resolution catalogue of radio
  sources in this field \citep{Brienza2017}.

A second possibility is that this source is the continuation of the radio filament F3 or a diffuse synchrotron emission associated with an high-redshift galaxy cluster, since the closest galaxy in projection is the galaxy at z=0.637 detected at  $\approx$11$^{\prime\prime}$ from the source radio peak. In the latter case the radio largest linear size of the source would be about 570\,kpc, its radio power\footnote{We assumed a spectral index $\alpha=2$ as derived for the brightest region of the emission, see text above.} at 144\,MHz $\sim 1.7\times 10^{26}$\,W/Hz and at 1.410\,GHz $\sim 1.8\times 10^{24}$\,W/Hz, 
still consistent with typical radio powers observed for radio halos (see \citealt{Giovannini2009}, \citealt{Feretti2012} and \citealt{vanWeeren2019}).
Assuming the scaling relation $P_{\rm 150\,MHz}$-$M_{\rm SZ, 500}$, $P_{\rm 1.4\,GHz}$-$M_{\rm X, 500}$, and $P_{\rm 1.4\,GHz}$-$L_{\rm X, 0.1-2.4\,keV}$ \citep{vanWeeren2020,Yuan2015}, we predict a mass of the system $M_{\rm 500}$ in the range $8-11\times 10^{14}M_{\odot}$ and an X-ray luminosity $L_{\rm X, 0.1-2.4\,keV}=4-18\times 10^{44}\,M_{\odot}$.  
These values are consistent with those observed for galaxy clusters at similar redshift \citep[e.g., ][]{Schneider2015}. However, we note that such a system would be very massive and therefore very rare.

Finally, a third possibility is that this source is an Odd Radio Circle \citep[ORC,][]{Norris2021}. The size, morphology and spectral properties resemble those of this new class of sources discovered with Australian Square Kilometre Array Pathfinder (ASKAP). The known ORCs have been observed in the direction of high-redshift galaxies and, interestingly, close to the location of patch P one galaxy is detected at z=0.637. However, these objects show a bright limb not observed here and the galactic latitude of patch P is lower than that observed for ORCs. If confirmed, this would be the first ORC seen by LOFAR.

\subsection{Spectral index profile}

In Fig.\,\ref{A523_spix_profile}, the radial profiles of the radio
brightness at 144\,MHz (black dots) and 1.410\,GHz (blue triangles)
are shown in panel (a), while the corresponding spectral index
profile is shown in panel (b). These profiles are derived by
using concentric annuli centered on the radio peak at 144\,MHz on the
same 65$^{\prime\prime}$ radio brightness images used to derive the
spectral index spatial distribution. The annuli have widths of one
beam size. The statistics are evaluated after masking all the embedded
discrete sources as shown in the right panel, including S1,
whose steep tails could otherwise contaminate our estimation. A
  few additional sources have been masked with respect to those
  labeled in Fig.\,\ref{hr} because they are characterized by a radio
  brightness peak $\gtrsim\,8\,\sigma$ at 1.410\,GHz.

\begin{figure}
\includegraphics[width=0.45\textwidth]{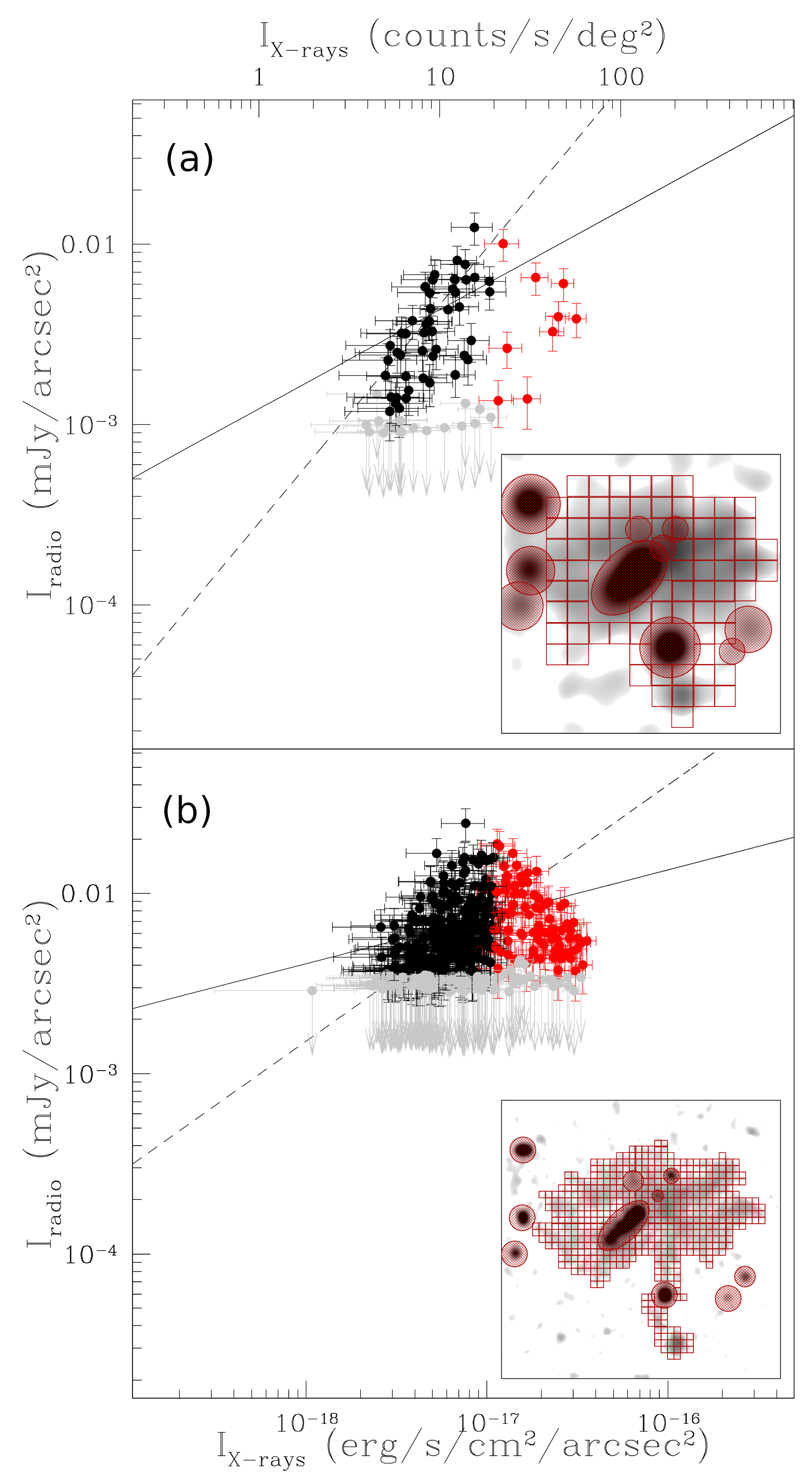}
\caption{Radio 144\,MHz versus X-ray surface brightness of the cluster at 65$^{\prime\prime}$ (panel (a)) and at 20$^{\prime\prime}$ (panel (b)): upper limits of the radio emission in gray, points corresponding to X-ray emission larger than 20\,counts/s/deg$^2$ in red and smaller than 20\,counts/s/deg$^2$ in black. The continuous line represents the power-law fit including all data points, while the dashed line represents the fit only considering points corresponding to an X-ray emission lower than 20\,counts/s/deg$^2$ (including upper limits). The inset panels show the grid on the radio emission of the cluster at 144\,MHz.}
\label{A523_corr}
\end{figure} 

Panel (c) in Fig.\,\ref{A523_spix_profile} shows the pixel
distribution of the spectral index image at 65$^{\prime\prime}$ (top
left panel in Fig.\,\ref{A523_spix}), as a function of the distance
from the radio peak at 144\,MHz, after masking embedded discrete
sources.  The pixel distribution indicates that the low-resolution
image only properly samples regions out to 500-600\,kpc due to the
sensitivity cut we applied. In this region, the radial profile
provides an average spectral index $\alpha\sim 1.2$, in agreement
within the uncertainty with the mean over the spectral index
image, see \S\,\ref{Spectral_index_images}.  Since we do not apply any
sensitivity cut to derive the spectral index radial profile, the
radial profile can give us some indication about the spectral index
behaviour also beyond 600\,kpc, in regions where the radio brightness
is below the 3$\sigma$ level.  In this way, possible flattening
and/or steepening in the spectral index distribution can be possibly
detected. Beyond 600\,kpc, we derive a lower limit of the spectral
index of $\alpha\,\gtrsim\,$1.5. Results in regions covered by the
most external annulus must be interpreted with caution. Indeed, the emission here is below the noise level of an individual beam and, as we used a deconvolution threshold of 1$\sigma$ and therefore signal below this threshold it is not deconvolved, LOFAR emission may result in a brighter signal than reality, leading to artificial spectral steepening.  Moreover, we note that even though we match VLA and LOFAR \emph{uv}-range, the density of the points in the inner of the \emph{uv}-plane is much higher in LOFAR than in the VLA data. However, the impact of this on the images is very difficult to estimate.   The diffuse emission in A523 does not show a circular shape and the radio and
  X-ray emission do not fully overlap. A substantial offset
  between the peak in X-ray surface brightness and radio brightness is
  also present at 144\,MHz, the largest absolute offset observed for
  radio halos and among the largest fractional offsets (offset/size)
  when giant radio halos only are considered \citep{Feretti2012}, as
  expected in case of a magnetic field fluctuating on large spatial
  scales \citep[e.g., ][]{Vacca2010}. This offset makes the
radial profile of the spectral index and its link with the thermal
properties of the system difficult to interpret. No steepening is
  observed even when the spectral index is computed as a function of
  the distance from the peak in the X-ray emission (not shown here).

\subsection{Spectral index of discrete sources}
\label{spixds}
The presence of the radio galaxies identified in \S\,\ref{lowfreq} is apparent in the spectral index images shown in Fig.\,\ref{A523_spix}. Flux densities of these sources have been derived as described in \S\,\ref{lowfreq}.
Spectral index values between 144\,MHz 1.410\,GHz are given in 
Table\,\ref{spix_ps}.  
For the sake of comparison, we report in Table\,\ref{spix_ps} the 
  flux density and spectral index values by \cite{deGasperin2018},
derived by using observations at 147\,MHz from the TIFR GMRT Sky
Survey \citep[TGSS, ][]{Intema2017} and at 1.4\,GHz from the NVSS
\citep{Condon1998}. The sources in the catalog have been identified by
imposing a distance in RA and Dec respectively of $\lesssim
25^{\prime\prime}$ from our sources, comparable to the TGSS spatial
resolution.  TGSS flux densities are available for four out of
  ten sources and the NVSS flux densities for six of them.  The
remaining sources are probably too faint to be detected at the
corresponding sensitivity of TGSS and/or NVSS. The NVSS fux
  density of the sources agrees with our VLA values within at
  most 1.3$\sigma$, while the TGSS measurements agree with our LOFAR
  flux densities within at most 1.9$\sigma$, with LOFAR measurements
  being systematically slightly higher, \footnote{\label{foot} The
    displacement $\sigma$ has been computed as $|S_{\rm XX}-S_{\rm
      YY}|/\sqrt{E_{\rm XX}^2+E_{\rm YY}^2}$, where $E_{\rm XX}$ and
    $E_{\rm YY}$ are the uncertainties in $S_{\rm XX}$ and $S_{\rm
      YY}$, as given in Table\,\ref{spix_ps}.}. This is likely due to
a difference in the flux density scale of our LOFAR observations with
respect to the TGSS data that, comparing the flux densities of these
four sources, amounts to a factor of about $1.4-1.6$. 
  Position-dependent flux density scale variations have been found in
  the TGSS by other authors \citep{Hurley-Walker2017}. The
  \cite{deGasperin2018} catalog reports spectral index information for
  six of our sources (S1, S5, S6, S7, S8, and S9, Col.\,10,
  Table\,\ref{spix_ps}), derived from the TGSS and NVSS flux density
  values. However, for two of them (S6 and S9), the TGSS flux density
  given in the catalog is zero (see Col.\,6, Table\,\ref{spix_ps}),
  therefore it is unclear how the spectral index has been derived and
  we will neglect them in the following. The spectral indices of the
  remaining four sources differ with respect to our by about
  2.3$\sigma$ for S1, 1.7$\sigma$ for S5, 0.4 $\sigma$ for S7, and
  1.5$\sigma$ for S8 (for each source $\sigma$ has been computed with
  the same approach described in Footnote\,\ref{foot}). Our flux
  density measurements for S1 include the contribution from the
  diffuse emission. Given the size of the source and the mean radio
  brightness of the overall diffuse emission, this term amounts to
  about 35\,mJy and 2\,mJy respectively at 144\,MHz and
  1.4\,GHz. However, as per our knowledge, the flux densities given in
  the public catalog include this contribution as well and therefore
  the difference in the flux densities and in the spectral indices can
  not be ascribed to it. Our spectral index for source S1 is flatter
  than the published value, likely also because of the slightly higher
  flux density measured from the NVSS with respect to our VLA
  measurement. For the sources S7 and S5, the flux densities at
  1.4\,GHz appear to be consistent within 2$\sigma$, therefore the
  difference in the spectral index is entirely due to our higher flux
  density at 144\,MHz.

\begin{table*}
\caption{Results from the fit of the radio 144\,MHz brightness versus X-ray emission for regions characterised by an X-ray counts/s/deg$^2$ given in Col.\,1. In Col.\,2, 3 we report $a$ and $lg(b)$, while in Col.\,4, 5 the Pearson $r_{\rm P}$ and Spearman $r_{\rm S}$ coefficients. In Col.\,6 the resolution of the radio image is given.}
        \begin{tabular}{cccccc}
         \hline
          \hline
     X-rays counts/s/deg$^2$ &$a$& $lg(b)$& $r_{\rm P}$ &$r_{\rm S}$ & Resolution ($^{\prime\prime}$)\\
          \hline
          \hline

0-10 &$0.93\pm0.16$&$13.91\pm 2.80$&0.36&0.29&20\\
0-20&$0.71\pm0.07$&$9.96\pm 1.21$&0.38&0.42 &20 \\         
0-30&$0.61\pm0.06$&$8.31\pm 0.95$&0.36&0.41&20\\
0-40&$0.48\pm0.05$&$6.09\pm 0.84$&0.36&0.41&20\\
0-50&$0.35\pm0.04$&$3.78\pm 0.76$&0.32&0.38&20\\
0-60&$0.28\pm0.04$&$2.66\pm 0.72$&0.28&0.36&20\\
0-70&$0.26\pm0.04$&$2.29\pm 0.70$&0.27&0.36&20\\
0-30&$1.06\pm0.16$&$15.89\pm 2.81$&0.41&0.45&65\\
0-70&$0.55\pm0.12$&$7.13\pm 2.12$&0.41&0.48&65\\

\hline
            \hline
        \end{tabular}
        \label{stat}

\end{table*}

\section{Comparison with cluster X-ray properties}
\label{Radio_versus_X}

In this section we compare the radio properties of the cluster with the mass, X-ray emission, temperature, pressure, and entropy, looking for correlations among them as they 
contain information on the thermodynamic history of the system \citep[see, e.g., ][]{Voit2005,Shi2020}. To this end, we used the X-ray images presented by \cite{Cova2019}.

\subsection{Local Radio  - X-ray correlation}

\cite{Cova2019} compared the X-ray emission of the system with the radio properties at 1.4\,GHz by \cite{Girardi2016}. Their results confirm that this radio halo is one of the most significant outliers with respect to the X-ray - radio correlation for radio halos (see also \citealt{Giovannini2011}). Moreover, they find that the X-ray and radio brightness at 1.4\,GHz do not show a point-to-point correlation. Here, we investigate the local radio - X-ray correlation using the data at 144\,MHz. 

In Fig.\,\ref{A523_corr}, we compare the X-ray brightness in the
0.5-2.5\,keV energy band with the radio brightness at 144\,MHz, by
using the images at 65$^{\prime\prime}$ (panel (a)) and at
20$^{\prime\prime}$ (panel (b)) spatial resolution. Compact sources
have been masked and the diffuse emission has been covered using
a grid with a cell-size equal to the beam FWHM, as shown in the
inset. We show points corresponding to an X-ray signal
$<20$\,counts/s/deg$^2$ in black, points corresponding to an X-ray
signal $>20$\,counts/s/deg$^2$ in red, with radio upper limits
in gray. We measured the radio versus X-ray surface brightness
  for each cell and, in order to deal with physical units, we
  converted it from counts/s/deg$^2$ to erg/s/cm$^2$/arcsec$^2$, using
  the neutral hydrogen column density derived from HI4PI Survey
  \citep{Nadia2016}, the XMM/MOS Thin Count rate conversion factor
  (this detector is indeed used as a reference for deriving the
  scaling factor of the other detectors), and an APEC model 
  mean temperature of 4.3\,keV and abundance of 0.2 the solar one obtained for A523 in \cite{Cova2019}.  Then we
  fitted in logarithmic scale a power-law of the form $I_{\rm
    radio} = b(I_{\rm X-rays})^a$ to the distribution, including all
the data, with the \emph{scipy.optimize} python package.  At
65$^{\prime\prime}$, we find $a=0.55\pm0.12$ and
  $lg(b)=(7.13\pm2.12)$ with a Pearson correlation coefficient
\citep{Pearson1895} $r_{\rm P}=0.41$ and a Spearman correlation
coefficient \citep{Spearman1904,Myers2003} $r_{\rm S}=0.48$, while at
20$^{\prime\prime}$ $a=0.26\pm0.04$ and $lg(b)=(2.29\pm0.70)$ with
$r_{\rm P}=0.27$ and $r_{\rm S}=0.36$.  Despite the
weak correlation, it is not so flat as at 1.4\,GHz
\citep[$a=0.095\pm0.08$, $r_{\rm S}$ = 0.27, $r_{\rm P}$ =
0.28, ][]{Cova2019}.

If we progressively exclude from the fit points at high X-ray
brightness, the correlation becomes steeper and steeper, and slightly
stronger.  The high resolution radio image provides better
  statistics and, in this case, the Pearson and Spearman coefficients
  reach both their maximum value ($r_{\rm P}=0.38$ and $r_{\rm
    S}=0.42$) for an X-ray signal $<20$\,counts/s/deg$^2$. When we
  select only points corresponding to an X-ray signal
  $<20$\,counts/s/deg$^2$, we find $a=0.71\pm0.07$ and
  $lg(b)=(9.96\pm1.21)$.  In Fig.\,\,\ref{A523_corr}, we show the two
fits: the continuous line includes all data points, while the dashed
line only points with an X-ray signal $<20$\,counts/s/deg$^2$. 
In both
cases, upper limits are taken into account. In Table\,\ref{stat}, we
report the values of $a$, $lg(b)$ by the fitting routine, and the
corresponding Pearson and Spearman coefficients when fitting points
corresponding to different ranges of X-ray counts/s/deg$^2$. At
65$^{\prime\prime}$ the statistics quality is poor but includes
regions of faint radio emission. At 20$^{\prime\prime}$ we have more
cells and we are able to better investigate the behaviour of bright
patches that are averaged down by fainter surrounding regions at lower
resolution. 
Results obtained excluding strong X-ray emission regions,
typically coinciding with the cluster center, indicate that two trends
are present: one flatter including all data points and one steeper
including only faint X-ray emission.  For the radio halo in A2255,
\cite{Botteon2020a} showed that higher thresholds in $I_{\rm radio}$
translate to flatter slopes of the $I_{\rm radio}$- $I_{\rm
  X-rays}$. Close to the threshold, the selection tends to pick up
only the up-scattered values, with a consequent underestimate of the
slope of the correlation in faint X-rays regions. Here, we are
possibly observing a similar behaviour, when applying a selection on
the X-rays instead than in the radio brightness. However, in our case,
two trends can be qualitatively distinguished in the plots in
Fig.\,\ref{A523_corr} (black and red points) even without applying any
cut.

\begin{figure*}
\includegraphics[width=1.0\textwidth]{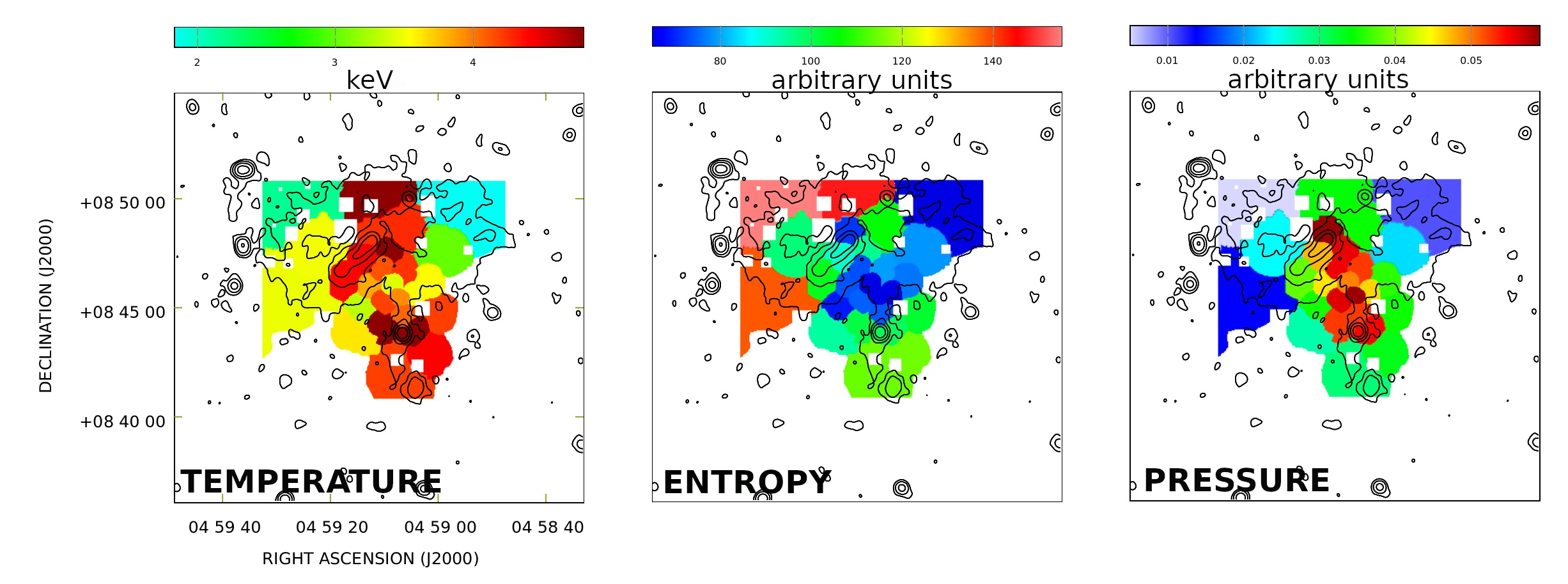}
\caption{Temperature (\emph{left panel}), entropy (\emph{middle panel}), and pressure (\emph{right panel}) images in colors, taken from \protect\cite{Cova2019}. While temperature is shown in keV, entropy and pressure are shown in arbitrary units. Contours represent the radio brightness at 144\,MHz at 20$^{\prime\prime}$, start at 3$\sigma$ ($\sigma=0.4\,$mJy/beam) and increase by a factor 4. They have been derived from the LOFAR images presented in this work. The field of view is the same in all the images.}
\label{thvsbr}
\end{figure*}

The behaviour observed in A523 differs from what observed for
  other clusters, i.e., A520 \citep{Govoni2001b,Hoang2019} and
  1E\,0657-55.8 \citep{Shimwell2014}, which appear to be characterized
  by respectively an almost flat correlation or by not significant
  correlation at all. Indeed, a steeper sub-linear correlation
  emerges when brighter X-ray regions are excluded from statistics,
  although this correlation is weak.  The behaviour observed in this
  system could reflect the complexity of the dynamical state of the
  system. The cluster is undergoing a primary merger along the SSW-NNE
  direction \citep{Girardi2016,Golovich2019} and likely a secondary
  merger along the ESE-WNW direction \citep{Cova2019}.  The flat
  component, weaker in radio, could be related to the primary merger
  brighter in X-rays. Our LOFAR data show a faint emission emerging in
  the north-east and in the south of the system, completely buried by
  the noise at higher frequencies. The steep component corresponds to
  regions stronger in radio and less bright in X-rays, being likely
  dominated by the emission in the north of the system. This structure
  is co-spatial with the possible secondary merger and the higher
  radio brightness associated with it could be the result of the
  superposition of the two mergers at this location.  While the radio
  plasma along the SSW-NNE direction is accelerated only by the
  primary merger and only visible at LOFAR frequencies likely due to
  ageing, the radio plasma in the perpendicular direction gains energy
  also through the secondary merger, likely explaining the observed higher
  radio brightness.

\subsection{Radio emission and cluster mass}
The mass of clusters hosting radio halos correlates with the radio-halo power at 1.4\,GHz \citep[e.g., ][]{Yuan2015}, and at 150\,MHz \citep[e.g., ][]{vanWeeren2020}. Numerical simulations show that the SZ effect is an excellent proxy of the cluster mass, with an intrinsic scatter of $\lesssim 10\%$ \citep[e.g.,][]{Nagai2006}, even if recent observations found a scatter up to $\sim$13\% that can be significantly reduced when non-thermal pressure associated with ICM turbulent motions are taken into account \citep[e.g.,][]{Yu2015}. Galaxy cluster masses derived from X-ray observations can be affected by the dynamical state of the system. The cluster might be in the adiabatic expansion phase and, in this case, X-ray proxies such as luminosity and temperature could be lower than during the initial state, causing an underestimate of the mass of the system \citep[see e.g., ][]{Ricker2001} with a bias up to $\sim$30\% (e.g., \citealt{Nagai2007}).
No SZ mass estimate is available for A523 and the X-ray derived mass is $M_{\rm X, 500}=2.2-3.6\times 10^{14}\,M_{\odot}$.
Considering our flux measurements at 1.410\,GHz and 144\,MHz, taking into account the different cosmology, the frequency scaling when appropriate, and k-correction\footnote{A spectral index $\alpha=1.2\,\pm\,0.2$ has been assumed, please refer to \S\ref{Spectral_index_images}.}, the value predicted according to the  $P_{\rm 1.4\,GHz}-M_{\rm X, 500}$ scaling relation in \cite{Yuan2015} is $M_{\rm 500}\,\approx\,6-10\,\times\,10^{14}\,M_{\odot}$, while
$M_{\rm 500}\,\approx\,6-9\,\times\,10^{14}\,M_{\odot}$ from the best fit relations $P_{\rm 150\,GHz}-M_{\rm SZ, 500}$ by \cite{vanWeeren2020}.

The values derived are larger than the mass obtained from X-ray
observations. Concerning the estimate from 144\,MHz data and the
\cite{vanWeeren2020} scaling relation, we stress that the
scaling relation is based on SZ data, while the mass estimate we are
using here is derived from X-rays. If we assume that the observed
X-ray mass is underestimated by about 30\%, we expect a mass of the
system of $M_{\rm 500}\,\sim\,5\,\times\,10^{14}\,M_{\odot}$, still
lower than the value obtained from the scaling relation. In all these
derivations the uncertainties both in the scaling relations and in our
radio powers have been taken into account.

Based on $\sigma_{\rm v}$-mass scaling relations from optical
  data, \cite{Girardi2016} derive a cluster mass $M_{\rm
    500}=5-6\times 10^{14}\,M_{\odot}$ that is in better agreement
  with the values predicted by the radio power of
the diffuse emission if this is a radio halo. 
We note that the uncertainty of the optical mass estimate varies between 7\% and 30\%.
 
  Overall, we find that the source remains outside the observed
  correlation both at 1.4\,GHz and 144\,MHz. However, we note that
the scaling relation is not well constrained in the mass range of
A523, because most of present studies rely on high mass clusters. We
need therefore to extrapolate the correlation in a poorly sampled
region of the radio brightness-mass diagram. Moreover, the
emission we observe in A523 could be the result of a superposition of
different sources (diffuse emission associated with the two
  merger processes, but also the patch P and the filaments F1, F2 and
  F3, which may not belong to the radio halo). An estimation of the
flux density associated with the filaments is very difficult because these
structures are embedded in the diffuse large scale emission.

\subsection{Radio brightness versus temperature, entropy and pressure}

We compared the radio brightness distribution at 144\,MHz with the temperature $T$, pressure $P$ and pseudo-entropy $s$ images of the thermal gas by \cite{Cova2019}. 
As stated in that paper, temperatures $T$ are directly derived through the spectral fitting, while the pressure $P$ and the entropy $s$ are calculated following \cite{Rossetti2007}
\begin{equation}
P=T \times {\rm EM}^{1/2}\,{\rm keV\,cm^{-5/2}\,arcmin^{-1}}
\end{equation}
and
\begin{equation}
s=T \times {\rm EM}^{-1/3}\,{\rm keV\,cm^{5/3}\,arcmin^{-2/3}},
\end{equation}
where EM is the projected emission measure. Both pressure and entropy values are pseudo quantities, meaning that they are projected along the line of sight.

In Fig.\,\ref{thvsbr} we show the overlay between the radio brightness
at 144\,MHz at 20$^{\prime\prime}$ and the temperature (first column),
the entropy (second column), and the pressure (third
column). Temperature and entropy decrease going from the south of the
cluster towards the center, and increase again moving further to the
north. The radio emission mainly sits in the north of the cluster
starting from the location where temperature and entropy increase
again. A low entropy strip with values $<80$ (arbitrary units)
cuts the cluster in two parts, one hosting bright radio emission and
one with very little radio emission. This strip has the same spatial
location as the filament F2 and as part of F1. The pressure image is
characterized by an elongation in the SSW-NNE, following the merger
axis, with larger pressure values along the filament F3. Moreover,
an enhancement in the temperature, entropy and pressure images
can be identified corresponding to patch P, indicating that this
region is dynamically active and therefore favours a
  possible interpretation as a phoenix source or a revived fossil plasma.

Overall, the bulk of the radio emission is perpendicular to the main
elongation in temperature, entropy and pressure, even if a minor
elongation of these three thermodynamic quantities can be identified
in the same direction of the brightest regions in radio.  
  Moreover, a hint of a pressure gradient seems to be present at the
  location of the tail of source S1. Available data and the current
  analysis are not suitable to discriminate whether this could, or
  could not, be linked to the diffuse emission on large scales observed
  in this system.  Further analysis is left to follow-up works.

\begin{figure}
\includegraphics[width=0.45\textwidth]{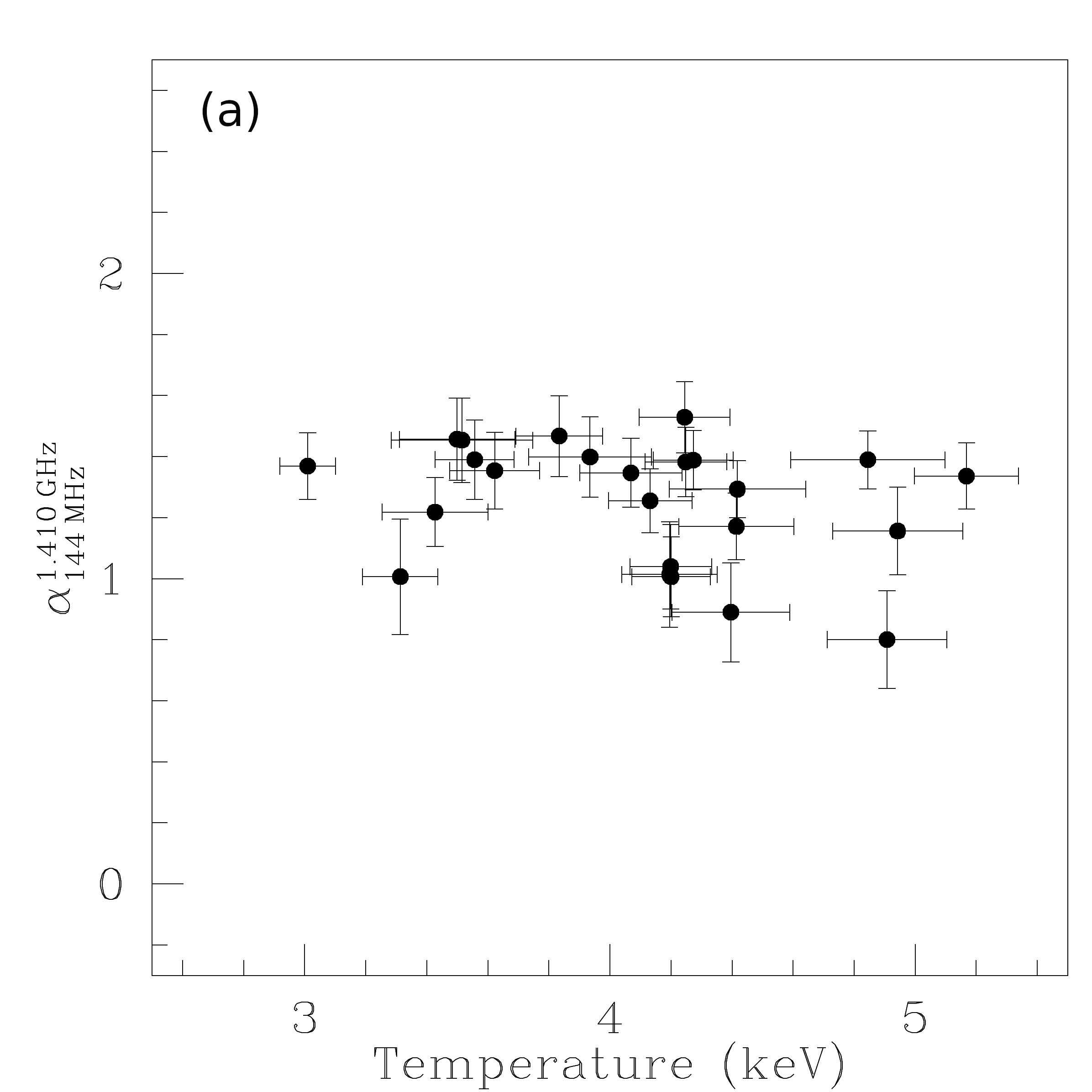}
\includegraphics[width=0.45\textwidth]{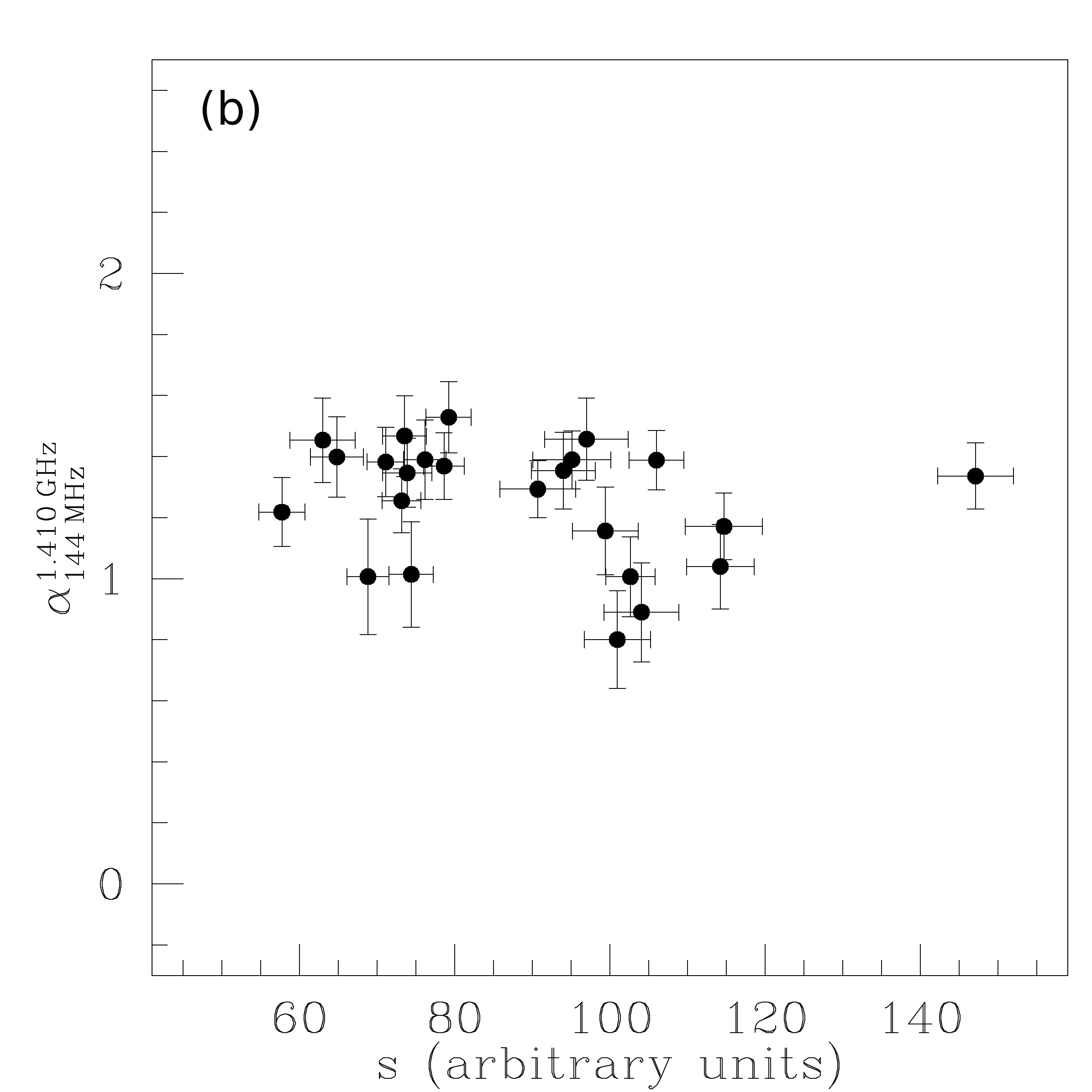}
\caption{\emph{Panel (a):} Spectral index between 1.410\,GHz and 144\,MHz versus thermal gas temperature of the cluster. \emph{Panel (b):} Spectral index between 1.410\,GHz and 144\,MHz versus thermal gas pseudo-entropy of the cluster. These plots have been obtained with the low-resolution spectral index images averaged in the same regions of the thermodynamic maps. The temperature and pseudo-entropy values are from \protect\cite{Cova2019}.}
\label{A523_spix_vs_T_XMM}
\end{figure} 

\subsection{Spectral index versus temperature and entropy}

Mergers can boost the X-ray luminosity and temperature of a cluster
\citep{Ricker2001,ZuHone2011}. If a fraction of the gravitational
energy dissipated during the merger is available to re-accelerate
radio emitting particles, high temperature and entropy regions may be
expected to have a flatter spectrum.  The study of any
  correlation between thermodynamic X-ray quantities and spectral
  index of diffuse radio emission in galaxy clusters has been
  conducted only for a limited number of clusters and present results
  are controversial.  An anti-correlation between spectral index and
temperature is indicated by the results reported by
\cite{Feretti2012}, who analyzed clusters with $T<8$\,keV,
$8<T<10$\,keV and $T>10$\,keV and derived a trend of decreasing
average spectral index for the three samples. A detailed
  investigation reveals this effect in the cluster A2744 in regions
of very different temperatures, spanning from about 5\,keV up to more
than 10\,keV \citep{Orru2007}. However, \cite{Pearce2017} in
A2744 and \cite{Shimwell2014} in the 1E\,0657-55.8 cluster do not find
significant evidence of this link. Finally, a possible evidence of
  anti-correlation between spectral index and pseudo-entropy in A2255 has been found by \cite{Botteon2020a}. In order to better
  understand the presence of a possible anti-correlation between
  spectral index and thermodynamic quantities in galaxy clusters, we
  need to investigate a larger statistical sample.

In Fig.\,\ref{A523_spix_vs_T_XMM}, we plot temperature (panel (a)) and  pseudo-entropy (panel (b)) against the spectral index. In these plots, the spectral index and the thermodynamic quantities have been averaged in the same regions shown in Fig.\,6 in \cite{Cova2019}. These plots do not show any clear anti-correlation. The Pearson and Spearman correlation coefficients are $r_{\rm P}=-0.28$ and $r_{\rm S}=-0.30$, and $r_{\rm P}=-0.26$ and $r_{\rm S}=-0.33$, respectively for the two plots.
Therefore, if any relation is present, it is likely very weak.

The lack of any link may be due to the fact that entropy is a tracer not only of merger shocks, but also of physical processes occurring in the ICM, as heating/cooling processes. Moreover, in this cluster, the temperature and the entropy show relatively small  variations, with values between 3 and 5\,keV and 50 and 150 (arbitrary units), compared e.g. to A2744 and A2255, therefore the lack of relations between spectral index and temperature and entropy is not inconsistent with previous findings.

\section{Discussion}
\label{discussion}

Historically the most prominent diffuse source in A523 has been classified as a radio halo \citep{Giovannini2011}. Subsequent observations revealed some peculiarities \citep{Girardi2016,Cova2019}: 
\begin{enumerate}
    \item the radio peak of the source is displaced with respect to the X-ray peak of the system;
    \item the source shows filaments of polarized emission up to 15-20\%;
    \item the radio emission at 1.4\,GHz and 144\,MHz shows a weak local correlation with the X-rays in the energy range 0.5-2.5\,keV;
    \item the radio emission at 1.4\,GHz does not follow the correlation with the global luminosity $L_{\rm X, 0.1-2.4\,keV}$ observed for other radio halos;
    \item the radio emission at 1.4\,GHz and 144\,MHz does not follow the correlation with the mass $M_{\rm 500}$ observed for other radio halos.
\end{enumerate}

In the following we discuss three possible scenarios for the origin of this radio source in light of our results.

\emph{Scenario 1.} The diffuse emission in A523 could be due to turbulence associated with the complex dynamical state of the system caused by {the main and a possible secondary merger}. 
The local radio-X-ray comparison at 144\,MHz suggests that the
  primary merger powers diffuse radio emission along the SSW-NNE
  direction, as shown  by the LOFAR observations, while the secondary merger further energizes particles along ESE-WNW, characterized by higher radio brightness than emission in the north-east and in the south-east and clearly detected both at 144\,MHz and at 1.4\,GHz.

\emph{Scenario 2.} 
The observed radio emission could be the revival of fossil plasma seeded by the central AGN S1 and later re-accelerated by the merger. In this case the correlations observed for radio halos are not expected. LOFAR images reveal several examples of systems where the AGN plasma has spread cosmic ray electrons over large areas (several hundreds of kpc), as recently observed by \cite{Brienza2021}, especially in presence of merger turbulence and shocks  \citep[e.g.,][]{Mandal2020}.
However, revived fossil plasma sources and phoenixes typically have  a
size of 300–400\,kpc and steep spectral indices ($\alpha\gtrsim 1.5$,
see \citealt{Mandal2020} as well as \citealt{vanWeeren2019} for a
review), different from what is observed in A523.

\emph{Scenario 3.}  The lack of an overall radio-X correlation
  could suggest that radio and the thermal plasma do not occupy the
  same volume, with the radio emission being a relic seen in
  projection, as proposed first by \cite{vanWeeren2011}.  A small
  inclination of the merging axis with respect to the line of sight
  ($\lesssim$10-20$^{\circ}$) could explain why a shock wave and a
  spectral gradient have been not observed. In cosmological
  simulations, face-on relics show complex morphologies that consist
  of filaments, possibly polarized, similar to the diffuse source in
  A523 \citep[e.g., ][]{Skillmann2013,Wittor2019,Wittor2021}. In these
  simulations, the spectral indices vary across the relics' surfaces,
  but they lack the typical spectral steepening towards the cluster
  center. However, this scenario is not supported by redshift data since there is no evidence for a merger
along the line of sight  \citep{Girardi2016,Golovich2019}.  Alternatively, an
  undetected component along the line of sight could be present if the
  merger is at the turnaround point or if the possible secondary
  merger is not entirely in the plane of the sky.

\section{Conclusions}
\label{conclusions}
In this paper we studied the properties of the diffuse emission in A523 by using new LOFAR observations at 144\,MHz and new VLA data at 1.410 and 1.782\,GHz. 
Our finding can be summarized as follows.

The new radio data reveal an unprecedented amount of detail about
  the properties of the source at radio wavelengths. The emission at
  144\,MHz appears more extended than at 1.4\,GHz, with a total flux
  density $S_{\rm 144\,MHz}=(1.52\pm0.31)$\,Jy and a size of about
  15$^{\prime}$ (i.e., 1.8\,Mpc). The source is characterized by a
  complex morphology consisting of three bright filaments, two in the
  north of the system already known from previous observations at
  1.4\,GHz, and a third one developing from the bright northern
  structure to the south of the cluster. The brightest region of the
  emission is elongated in the ESE-WNW direction also at
  144\,MHz. Thanks to the new LOFAR data, we detect for the first time
  additional regions of faint emission along SSW-NNE and a bright
  diffuse synchrotron patch in the south, characterized by a steep
  spectral index and of unclear origin. A local comparison of the
  radio and X-ray signal suggests the presence of two components, one
  brighter in X-ray and less bright in radio and the second the other
  way around. The new LOFAR and VLA data permit, for the first
  time, an investigation of the spectral properties of the source.
  Globally, we derive an average spectral index $\alpha_{\rm
    144\,MHz}^{\rm 1.410\,GHz}\,=\,1.2\,\pm\,0.1$ with a spectral
  steepening moving towards higher frequency. The spectral index does
  not show radial steepening but rather a complex spatial
  distribution.

Overall, our findings suggest that we are observing the overlapping of different structures, powered by the turbulence associated with the primary and a possible secondary merger. Our results and the current optical data do not support a relic interpretation of the source as a whole, while the relic nature of the northern filaments as well as the revived fossil plasma scenario can not be excluded.
Although we have derived valuable information about this complex source from the new radio data, additional optical and X-rays observations are necessary to better understand the geometry of the merger and to verify or exclude the presence of possible components along the line of sight. 

\section*{Acknowledgements}
\addcontentsline{toc}{section}{Acknowledgements}
%
This paper is in memory of our shiny meteor Francesco and of Luciano.
We thank the referee whose comments and suggestions helped to strongly improve the presentation of our results. This paper is based (in part) on data obtained with the International LOFAR Telescope (ILT) under project code LC10\_024. LOFAR (van Haarlem et al. 2013) is the Low Frequency Array designed and constructed by ASTRON. It has observing, data processing, and data storage facilities in several countries, that are owned by various parties (each with their own funding sources), and that are collectively operated by the ILT foundation under a joint scientific policy. The ILT resources have benefitted from the following recent major funding sources: CNRS-INSU, Observatoire de Paris and Universit\'e d'Orl\'eans, France; BMBF, MIWF-NRW, MPG, Germany; Science Foundation Ireland (SFI), Department of Business, Enterprise and Innovation (DBEI), Ireland; NWO, The Netherlands; The Science and Technology Facilities Council, UK; Ministry of Science and Higher Education, Poland. The present project is carried out within the framework of the LOFAR Magnetism Key Science Project \url{https://lofar-mksp.org/}. We acknowledge the computing centre of INAF - Osservatorio Astrofisico di Catania, under the coordination of the WG-DATI of LOFAR-IT project, for the availability of computing resources and support. VV, MM and GB acknowledge support from INAF mainstream project “Galaxy Clusters Science with LOFAR” 1.05.01.86.05. FL acknowledges financial support from the Italian Minister for Research and Education (MIUR), project FARE, project code R16PR59747, project name "FORNAX-B". FL acknowledges financial support from the Italian Ministry
of University and Research - Project Proposal CIR01$\_$00010. RJvW acknowledges support from the ERC Starting Grant ClusterWeb 804208.  AB acknowledges support from the VIDI research programme with project number 639.042.729, which is financed by the Netherlands Organisation for Scientific Research (NWO). AB and DNH acknowledges support from the ERC through the grant ERC-Stg DRANOEL n. 714245. MB acknowledges support from the Deutsche Forschungsgemeinschaft under Germany’s Excellence Strategy - EXC 2121 "Quantum Universe" - 390833306. D.W. is funded by the Deutsche Forschungsgemeinschaft (DFG, German
Research Foundation) - 441694982.



\section*{Data availability}
The data underlying this article will be shared on reasonable request to the corresponding author.

\bibliographystyle{mnras}
\bibliography{mnras} 



\bsp	
\label{lastpage}
\end{document}